\documentclass[journal,comsoc]{IEEEtran}
\usepackage[T1]{fontenc}

\ifCLASSINFOpdf
  \usepackage[pdftex]{graphicx}
  \graphicspath{{figs_and_tables/},{bio_photos/}}
  \DeclareGraphicsExtensions{.pdf,.jpeg,.png,.jpg,.eps,.PNG,.bmp}
\else
\fi
\usepackage{amsmath}
\usepackage[cmintegrals]{newtxmath}
\ifCLASSOPTIONcompsoc
 \usepackage[caption=false,font=normalsize,labelfont=sf,textfont=sf]{subfig}
\else
 \usepackage[caption=false,font=footnotesize]{subfig}
\fi
\usepackage{url}
\usepackage{hyperref}

\usepackage[inline]{enumitem}

\begin{document}
%
\title{Understanding Crowd Behaviors in a Social Event by Passive WiFi Sensing and Data Mining}
%
%
%

\author{Yuren~Zhou,~\IEEEmembership{Member,~IEEE,}
        Billy~Pik~Lik~Lau,~\IEEEmembership{Student~Member,~IEEE,}
        Zann~Koh,
        Chau~Yuen,~\IEEEmembership{Senior~Member,~IEEE,}
        and~Benny~Kai~Kiat~Ng
\thanks{They authors are with Pillar of Engineering Product Development, Singapore University of Technology and Design, Singapore.}
\thanks{E-mail: yuren\_zhou@sutd.edu.sg and yurenchau@sutd.edu.sg}
\thanks{Copyright (c) 2020 IEEE. Personal use of this material is permitted. However, permission to use this material for any other purposes must be obtained from the IEEE by sending a request to pubs-permissions@ieee.org.}}

%
%

\markboth{Manuscript accepted by IEEE Internet of Things Journal}%
{Shell \MakeLowercase{ZHOU \textit{et al.}}: Understanding Crowd Behaviors in a Social Event by Passive WiFi Sensing and Data Mining}
%



\maketitle

\begin{abstract}
Understanding crowd behaviors in a large social event is crucial for event management. Passive WiFi sensing, by collecting WiFi probe requests sent from mobile devices, provides a better way to monitor crowds compared with people counters and cameras in terms of free interference, larger coverage, lower cost, and more information on people's movement.
In existing studies, however, not enough attention has been paid to the thorough analysis and mining of collected data. Especially, the power of machine learning has not been fully exploited.
In this paper, therefore, we propose a comprehensive data analysis framework to fully analyze the collected probe requests to extract three types of patterns related to crowd behaviors in a large social event, with the help of statistics, visualization, and unsupervised machine learning. First, trajectories of the mobile devices are extracted from probe requests and analyzed to reveal the spatial patterns of the crowds’ movement. Hierarchical agglomerative clustering is adopted to find the interconnections between different locations. Next, $k$-means and $k$-shape clustering algorithms are applied to extract temporal visiting patterns of the crowds by days and locations, respectively. Finally, by combining with time, trajectories are transformed into spatiotemporal patterns, which reveal how trajectory duration changes over the length and how the overall trends of crowd movement change over time. The proposed data analysis framework is fully demonstrated using real-world data collected in a large social event. Results show that one can extract comprehensive patterns from data collected by a network of passive WiFi sensors.
\end{abstract}


%
\IEEEpeerreviewmaketitle

\section{Introduction}
\label{sec_intro}
%
%
%
%

\IEEEPARstart{M}{onitoring} and understanding crowd behaviors (visiting and moving) in large social events is an important task, as it can provide insights to event organizers for better event management and alerts of emergencies. Although traditional sensors, such as people counters and cameras, can monitor the number of people, they suffer from issues like coverage, cost, and lack of information about people's movement~\cite{kurkcu2017estimating}. With the quick expansion of mobile device ownership, new data sources become available for studies on understanding crowd behaviors (such as presence and movement) in large social events, mainly including GPS data, Bluetooth detection records, and WiFi probe request records.

Among the above three types of data, GPS data provide the most precise and accurate data regarding the movement of people. However, it has the problem of failure in an indoor environment, and it requires the development of a specially designed mobile application and the active participation of people. For example, in~\cite{wirz2013probing}, Wirz et al. conducted a real-world data collection for the Lord Mayor's Show 2011 in London to estimate the crowd density. They were able to cover a large area in London with the collected data, but the data were only from 828 pedestrians out of around half a million spectators in the event. In~\cite{blanke2014capturing}, Blanke et al. developed an official mobile application for a large Swiss event with carefully designed incentive systems, and finally, 28,000 users contributed their data, which was not many given the scale of the event.

Unlike GPS data, which are collected in a participatory manner, Bluetooth detection records and WiFi probe request records can be collected passively by sensing signals sent by the mobile devices over the air, which does not require any actions from event participants~\cite{zhou2018understanding}.
Therefore, they can save manpower costs on data collection and provide a much larger coverage on the entire population of participants. Moreover, while GPS can only be used outdoors, Bluetooth and WiFi technologies can be used to monitor crowds in both indoor and outdoor environments. Indoor examples include museums studied in~\cite{yoshimura2014analysis,casolla2019exploring,hong2018crowdprobe} and hospitals studied in~\cite{booth2019toward,ruiz2014analysis}. Outdoor examples can be found in~\cite{yoshimura2017analysis,wu2018crowdestimator} (pedestrian commercial districts) and~\cite{martinez2019evaluation,musa2012tracking} (city streets).
As a result, Bluetooth detection records and WiFi probe request records are more suitable for monitoring and understanding crowds in a large social event.

Compared with WiFi probe request records, Bluetooth detection records are more accurate due to a smaller range of Bluetooth~\cite{chilipirea2018identifying}. However, because it is more likely for a mobile device to have its WiFi turned on instead of its Bluetooth, Bluetooth detection records have a much smaller detection rate, which can be as low as 3\% of WiFi probe request records according to~\cite{schauer2014estimating}.
Sometimes, to increase the detection rate of Bluetooth scanning, special devices are distributed to event participants, such as tracking bracelets used in~\cite{jabbari2019ice}.
Moreover, instead of making Bluetooth sensors detect passing-by mobile devices, some studies deployed Bluetooth beacons in the studied area and made participants carrying special mobile devices that can collect signals sent by surrounding beacons~\cite{song2019using,dogan2019analyzing}. Such a sensing schema enables the precise localization of the participants and can provide real-time location-based services (such as guiding) to the participants~\cite{piccialli2019lessons,piccialli2019decision}.
However, similar to the GPS collection, the above two schemas of Bluetooth sensing require the distribution of special devices (or software) and active participation of people, which is not well scalable.
Therefore, to guarantee the detection rate and scalability in this work, WiFi probe request records are collected and analyzed to understand the crowd behaviors in a large social event.

WiFi probe requests are sent by mobile devices in a certain period to detect usable WiFi signals in the surroundings, and they contain information about the sending devices, such as their MAC (Media Access Control) addresses. The general communication range of WiFi is around 35 m for the indoor environment and more than 100 m for the outdoor environment. By combining received time of probe requests, MAC addresses they contain, and locations of the sensors, one can obtain location (movement) information of moving devices, and thus, of people. Although some recent mobile devices have started to use randomized local MAC addresses when they are not connected to WiFi for privacy concern, not all of them are equipped with such schema. According to a study conducted by Martin et al. in 2017~\cite{martin2017study}, devices with randomized MAC addresses take up less than 50\% of the entire sample, and there is no way to map probe requests to individuals with only MAC addresses. Also, it has been shown that, even with a dataset that contains a large portion of randomized MAC addresses, it is still possible to draw meaningful results regarding crowd behaviors~\cite{hong2018crowdprobe}.

Existing studies, which make use of WiFi probe requests to understand crowd behaviors in social events or given spaces, can be roughly divided into two categories based on whether the studied spaces are outdoor or indoor.
For outdoor environments, related studies are mainly focused on the development of a better WiFi-based passive sensing systems~\cite{lesani2018development,bonne2013wifipi}, estimation and analysis of crowd size and density~\cite{wu2018crowdestimator,weppner2016monitoring}, analysis of space utilization~\cite{prentow2015spatio}, precise localization of mobile devices~\cite{musa2012tracking}, and the extraction of trajectories of people~\cite{chilipirea2018identifying,traunmueller2018digital}.
For indoor environments, beyond ordinary visiting and moving statistics, more fine-grained results regarding crowd behaviors have been achieved, such as the social interactions between people studied in~\cite{hong2016socialprobe,shen2018snow}, because the spatial granularity of collected data is often higher in an indoor space due to bounded space and higher density of existing access points.
Although existing works have demonstrated the effectiveness of mining WiFi probe requests to understand crowd behaviors in both the outdoor and indoor areas, the thorough analysis of the data collected has not been done with enough focus.
Especially, approaches used to analyze the collected data in existing studies mainly include domain knowledge-based processing~\cite{ruiz2014analysis}, statistics~\cite{fukuzaki2014pedestrian,weppner2016monitoring}, and visualization~\cite{prentow2015spatio,traunmueller2018digital}. With the rapid development of machine learning algorithms, we believe that more unsupervised learning algorithms should be explored for mining WiFi probe requests. As a result, this paper is focused on filling this gap.

In this paper, a comprehensive data analysis framework is proposed to extract multi-aspect patterns regarding crowd behaviors from WiFi probe request records collected by passive WiFi sensor networks in a large social event. The framework consists of steps that are designed to preprocess and analyze the collected probe requests to extract spatial, temporal, and spatiotemporal patterns. First, MAC addresses that are not randomized are identified from the overall dataset, and their probe requests are preprocessed to form trajectories. The obtained trajectories are analyzed extensively to discuss the popularity of different locations and common patterns of people's movement. Hierarchical agglomerative clustering is adopted at this stage to extract interconnections between locations. Next, data of all detected MAC addresses are analyzed to extract temporal patterns about people's visits to the entire area and to different locations. $k$-means and $k$-shape clustering algorithms are applied to achieve clusters of days and clusters of locations, respectively. Finally, spatiotemporal patterns regarding people's movement are extracted and discussed.
To demonstrate the proposed data analysis framework, data collected from a large social event held in a popular tourist area in Singapore are used.

The contributions of this paper can be summarized as the following:
\begin{itemize}
  \item We propose a comprehensive framework for analyzing WiFi probe request records to extract spatial, temporal, and spatiotemporal patterns of crowd behaviors in a large social event.
  \item We apply and demonstrate the effectiveness of three different clustering algorithms, namely hierarchical agglomerative clustering, $k$-means clustering, and $k$-shape clustering, on extracting different hidden patterns from the probe request records.
  \item We demonstrate the proposed framework using real-world data collected in a large social event held in a popular tourist area.
\end{itemize}

We believe that the proposed data analysis framework along with the adopted clustering algorithms can benefit future related work as a comprehensive guide on analyzing data of WiFi probe requests. Meanwhile, the obtained results in the case study also provide useful insights to the research communities related to human mobility and tourism management.

\begin{figure}[!t]
\centering
\includegraphics[width=1.0\linewidth]{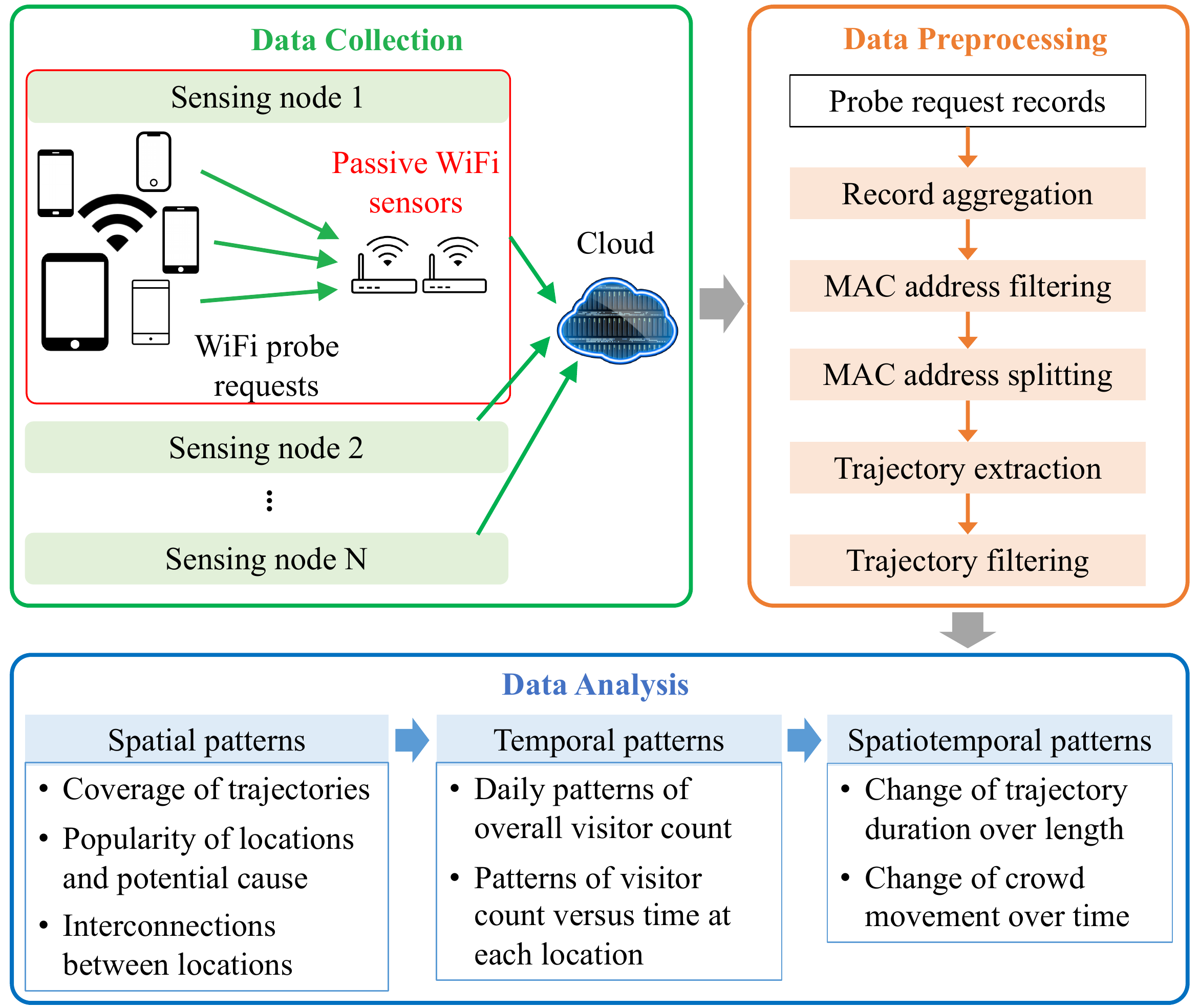}
\caption{Overall framework for understanding crowd behaviors through passive WiFi sensing and data mining.}
\label{fig_data_pipe}
\end{figure}

The overall framework of this study is summarized in Fig.~\ref{fig_data_pipe}. In the rest of this paper, Section~\ref{sec_data_preparation} describes details of data collection and preprocessing. The three aspects of data analysis, namely spatial patterns, temporal patterns, and spatiotemporal patterns are discussed in Section~\ref{sec_spatial}, Section~\ref{sec_temporal}, and Section~\ref{sec_spatiotemporal}, respectively. Finally, in Section~\ref{sec_conclusion}, we conclude this study and summarize the future work.

\section{Data Preparation}
\label{sec_data_preparation}

\subsection{Data Collection}

Passive WiFi sensing boxes are developed to collect probe requests from passing-by mobile devices in given locations. Each sensing box contains two sets of WiFi sniffers, and each sniffer is built on top of the Raspberry Pi Model B with additional WI-PI USB dongle for probe request collection. The WiFi sniffers have been encrypted to prevent any potential leakage of the collected data in case of theft or vandalism. To limit the volume of uploaded data, local processing of collected probe requests is conducted inside each sniffer to combine chronologically close probe requests, with a combining interval of three minutes. Processed probe request records are then uploaded through Long-Term Evolution (LTE) connection. Each record contains fields including MAC address, time of the first detection, time of the last detection, average received signal strength indication (RSSI), and the identifier (ID) of the collecting sniffer. To protect the privacy of people, no other information regarding the devices is reserved, and thus it is not possible to track back to individuals. More details regarding the hardware system can be found in~\cite{li2018experimental}.

\begin{figure}[!t]
\centering
\includegraphics[width=1.0\linewidth]{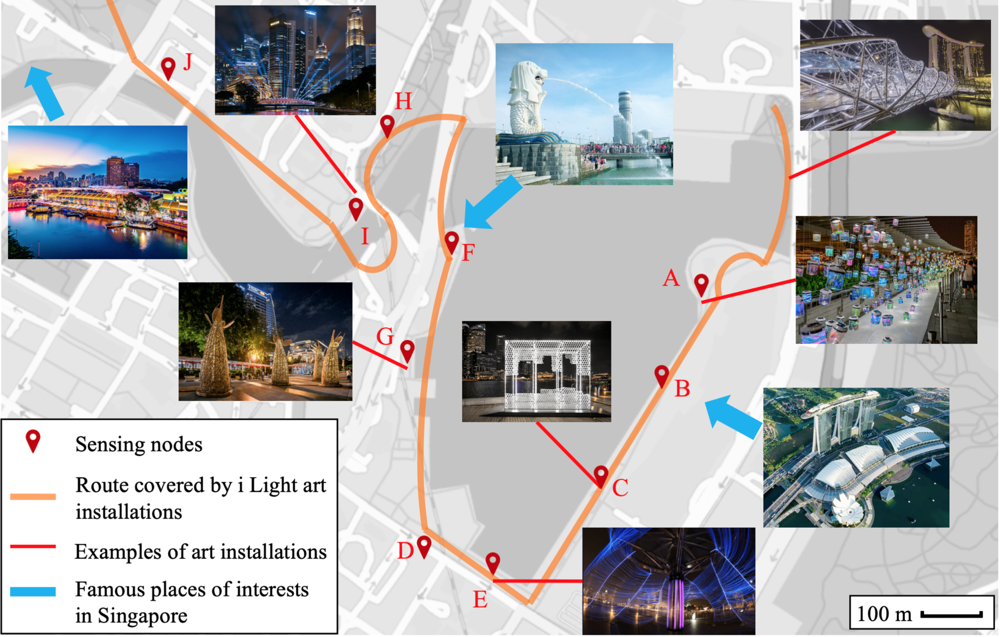}
\caption{Sensing nodes (locations of the deployed sensing boxes), examples of i Light art installations, and places of interests nearby the Marina Bay area in Singapore. The central darker grey area stands for water.}
\label{fig_deploy_map}
\end{figure}

Ten passive WiFi sensing boxes were deployed to collect information of passing-by crowds in ten different locations during a large outdoor social event, i Light Singapore, which was held from January 28th to February 24th 2019. i Light Singapore is Asia's leading sustainable light art festival held annually, and in 2019, 33 sustainable light art installations and five programming hubs were placed along Marina Bay, the Civic District, Singapore River, and Raffles Terrace at Fort Canning Park. Fig.~\ref{fig_deploy_map} shows the monitored outdoor locations A-J (referred to as the sensing nodes in the following text) along with related information of the event and the covered area. The route covered by the sensing nodes last around 2.5 $\mathrm{km}$ in total. Since the light art installations only operated in the nighttime, the data collection was conducted from 7 pm to 12 am daily. The valid data collection period started from January 29th and lasted till the end of the event. Overall, around 17 million probe request records were collected.

\subsection{Data Preprocessing}
\label{sec_data_preprocess}

Before data analysis, data preprocessing is conducted to filter out invalid data and transform the data into desired forms, following the steps depicted in Fig.~\ref{fig_data_pipe}.

First, probe request records are aggregated based on the sensing node with an "OR" logic. Since the two sniffers inside the same sensing node collect and upload data individually, there could be records with the same MAC addresses produced by these two sniffers at the same time. Since they both represent the same location of the same device, they are combined as one single record. After this stage, the field of sniffer ID in each probe request record is replaced by node ID.

\begin{figure}[!t]
\centering
\includegraphics[width=.6\linewidth]{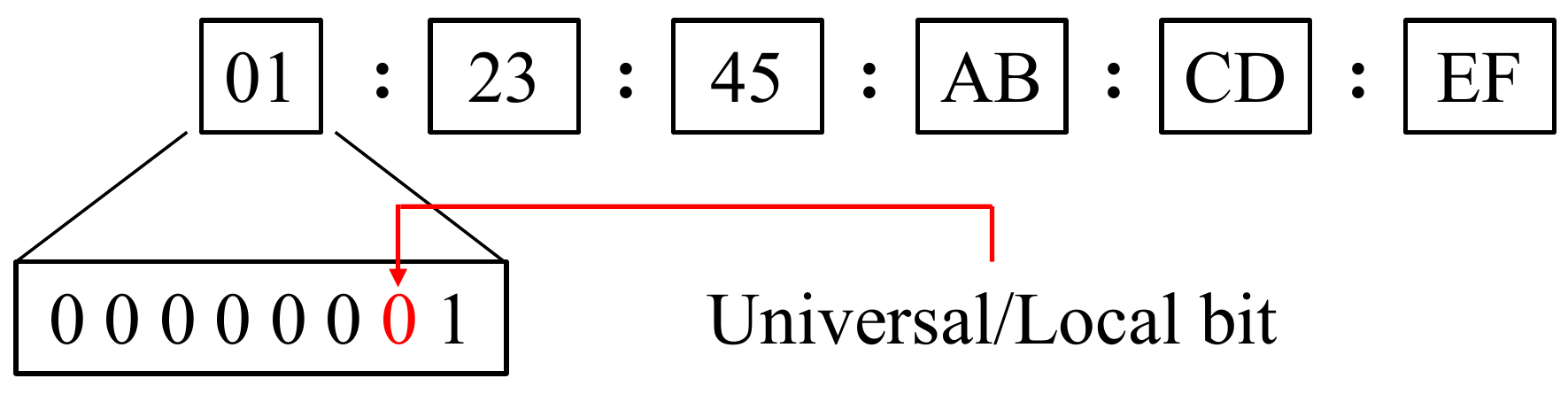}
\caption{Location of the Universal/Local bit in a MAC address.}
\label{fig_UL_bit}
\end{figure}

Second, unique MAC addresses are obtained from the probe request records. For each unique MAC address, the count of visiting days to the event area is computed, which is then used to filter out nearby static devices (e.g. surveillance cameras) and mobile devices of nearby workers. MAC addresses that appeared for more than four times a week on average are filtered out. After filtering, the dataset of the remaining probe request records is named as \textit{Dataset A}, which is later used to extract temporal patterns of the crowds in Section~\ref{sec_temporal}.

Next, the MAC addresses in \textit{Dataset A} are split based on whether they are globally unique or locally assigned. A globally unique (or burned-in) MAC address is the exact MAC address of each mobile device, which is a unique identifier assigned to its network interface controller by the manufacturer. Whereas, a locally assigned MAC address is assigned temporarily to override the global address. It can either be a MAC address randomized by the operating system for privacy considerations or generated for special local communication services such as mobile device-tethered hotspots~\cite{martin2017study}. When analyzing crowd movement, we remove the probe request records with local MAC addresses, since these addresses are temporary and not guaranteed to be unique. For global MAC addresses, in contrast, they are one-to-one matched to mobile devices and thus are used to extract trajectories and analyze the movement of the crowd in this study.
Whether a given MAC address is global or local can be identified by the Universal/Local (U/L) bit in it, which is the second-least-significant bit of the first octet of the address as shown in Fig.~\ref{fig_UL_bit}. When the U/L bit is set as 1, the MAC address is locally assigned~\cite{IEEE_MAC_guide}.

Finally, trajectories are extracted for global MAC addresses (i.e. visitors) on a daily basis. For each MAC address on each day when it appeared, all the probe request records are sorted by time of appearance first. Afterward, chronologically close records with the same node ID are combined as one node visit along the extracted trajectory. If the two combined records have an interval smaller than five minutes, the entire time span is considered as the staying time of the combined node visit. Otherwise, the interval between the two records is considered as the missing time of the device. Since the sensor networks were deployed in an outdoor environment and WiFi can transmit to a larger distance (more than 100 m), it is very likely that two sensing nodes collected the same probe request from the same mobile device. This issue can cause temporal conflicts between node visits along the trajectory. To solve such conflict, the following rules are applied:
\begin{enumerate}
  \item If one of the conflicting node visits has zero staying time and the other one does not, the one with zero staying time is abandoned.
  \item If both of the conflicting node visits have zero staying time, the one with a larger average RSSI is reserved and the other is abandoned.
  \item If both of the conflicting node visits have staying time larger than zero and one of the visits is entirely covered by another in time, the covered node visit is abandoned.
  \item If both of the conflicting node visits have staying time larger than zero and time periods when no conflict happened, the conflicting part is decided by the visit with a larger average RSSI.
\end{enumerate}
Moreover, since the sensing area is close to roads and streets with heavy traffic flows, mobile devices of people inside vehicles are likely to be detected as well. To remove such devices and only keep trajectories of pedestrians, the extracted trajectories that have zero staying time in total are filtered out.

After the entire preprocessing, the trajectories extracted from global MAC addresses can then be used to analyze the patterns related to the movement of crowds during the event. This particular dataset is named as \textit{Dataset B}.
By matching the three-byte prefix of the global MAC addresses in \textit{Dataset B} with the Organizationally Unique Identifier (OUI) lookup table from IEEE, vendor share of the included devices are computed, among which Samsung takes up around 36\%, Apple takes up 23\%, and other brands occupy the remaining 41\%. This observation roughly matches with the average mobile vendor market share from 2016 to 2018 in Singapore~\cite{VendorShare}. Based on the above, \textit{Dataset B} is a proper sample to investigate the patterns of the crowd movement.

\section{Spatial Patterns}
\label{sec_spatial}

As shown in Fig.~\ref{fig_data_pipe}, the extraction of spatial patterns is conducted to answer three important questions that could be of interest to the event organizers: 1) how the trajectories of people cover the event area, 2) how every node attracts people and what factor affects their popularity, 3) how people transit from one node to the other.
To answer the above questions, all trajectories in \textit{Dataset B} are exploited. Since the focus of this section is purely on the spatial patterns, all trajectories across all days of data collection are used to conduct the following analysis.

\subsection{Coverage of Trajectories}
\label{sec_com_traj}

\begin{table}[]
\renewcommand{\arraystretch}{1.1}
\centering
\caption{Split of trajectories based on number of nodes they visited and whether they are round trips or not.}
\label{tab_split_of_traj}
\resizebox{\linewidth}{!}{%
\begin{tabular}{c|c|c|cc|cc|cc|cc}
\hline
\begin{tabular}[c]{@{}c@{}}Number of\\ visited nodes\end{tabular} & 1 & 2 & \multicolumn{2}{c|}{3} & \multicolumn{2}{c|}{4} & \multicolumn{2}{c|}{5} & \multicolumn{2}{c}{\begin{tabular}[c]{@{}c@{}}6 and\\ more\end{tabular}} \\ \hline
Round trips & N/A & N/A & Y & N & Y & N & Y & N & Y & N \\
Split share (\%) & 63.17 & 19.81 & 3.07 & 6.07 & 0.31 & 3.56 & 0.45 & 1.60 & 0.23 & 1.74 \\ \hline
\end{tabular}%
}
\end{table}

\begin{figure}[!t]
\centering
\includegraphics[width=1.0\linewidth]{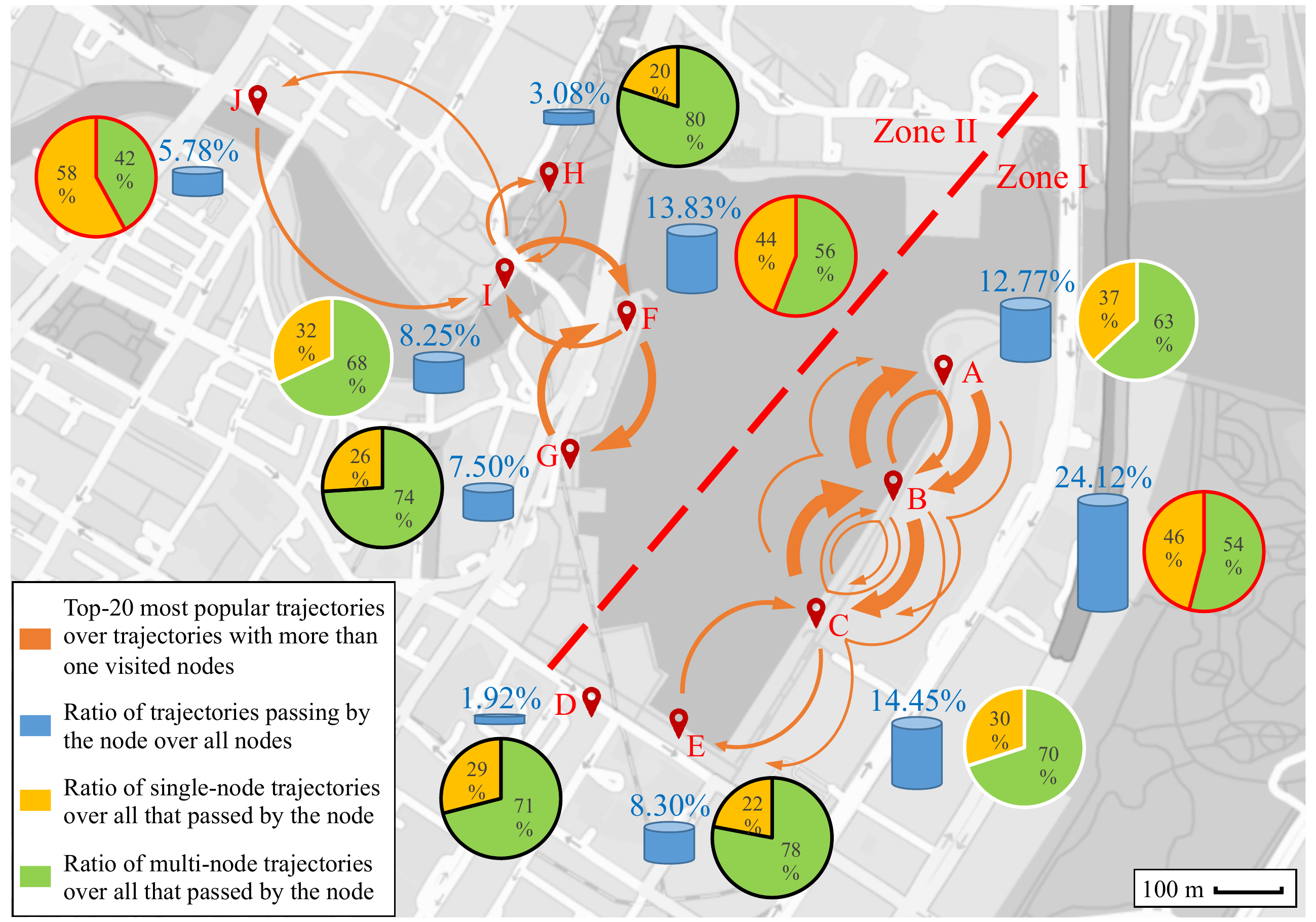}
\caption{Top-20 most popular trajectories taken by the visitors and statistics on popularity of each node. Ratios of trajectories passing by each node over all nodes sum up to one. For each node, the summation of ratio of passing by single-node trajectories and ratio of passing by multi-node trajectories equals to one.}
\label{fig_traj_cnt_and_popular_traj}
\end{figure}

\begin{figure}[!t]
\centering
\includegraphics[width=1.0\linewidth]{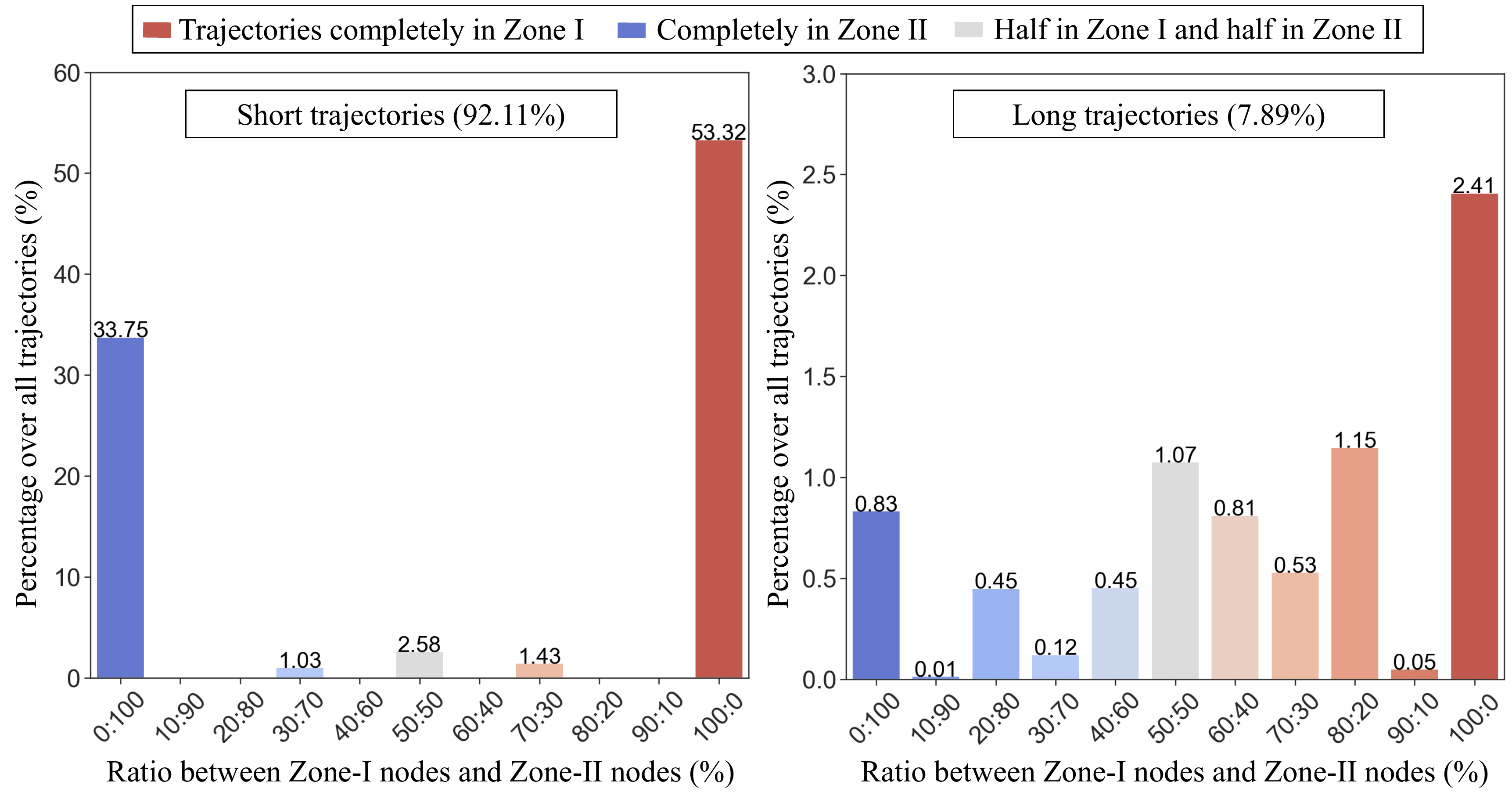}
\caption{Distribution of ratio between Zone-I nodes and Zone-II nodes for short trajectories (with maximum three nodes) and long trajectories (with more than three nodes). Percentages of all bars in the two figures sum up to one, representing all trajectories.}
\label{bar_plot_zone_split_ratio}
\end{figure}

In Table.~\ref{tab_split_of_traj}, count of the trajectories is split into different categories based on the number of nodes they visited and whether they are round trips (i.e. returning to the origin finally). From the figure, most (63.17\%) of the trajectories are single-node trajectories, which represent people who happened to pass by the node and did not continue to the next. Around 20\% of the trajectories last for two nodes. Among trajectories with three visited nodes, around one third are round trips. Whereas, for people who traveled for longer than three nodes, they tended not to return to their origins.

In Fig.~\ref{fig_traj_cnt_and_popular_traj}, the top-20 most popular trajectories among trajectories with more than one node are shown by orange arrows. From the figure, most of the popular trajectories last for two nodes and several have three nodes, and among the three-node trajectories, three out of five are round trips (e.g. B-A-B, B-C-B, C-B-C). Moreover, one can find that popular trajectories can be well separated into two groups: one group is located nearby nodes A, B, C, and E, and the other is located around node F, G, H, I, and J. Based on this grouping, the entire event area are separated into two zones, namely Zone I and Zone II as marked in the figure.

As shown by Table.~\ref{tab_split_of_traj} and Fig.~\ref{fig_traj_cnt_and_popular_traj}, the majority (92.11\%) of the trajectories have less than four visited nodes (termed as short trajectories) and each of the top-20 most popular trajectories only covers nodes either in Zone I or Zone II. To quantitatively describe the coverage of the trajectories in terms of Zone I and Zone II, Fig.~\ref{bar_plot_zone_split_ratio} shows the distribution of ratio between Zone-I nodes and Zone-II nodes for short trajectories (with maximum three nodes) and long trajectories (with more than three nodes). For visualization purposes, the ratio values are rounded into eleven levels as shown along the x-axis of the figure. From the figure, most of the short trajectories cover only one side of the event area, and among such trajectories, Zone I is more popular than Zone II. In contrast, for long trajectories (7.89\% of all trajectories), more than half (4.64\% among 7.89\%) have nodes both in Zone I and Zone II, and 23\% of such trajectories (1.07\% among 4.64\%) have half of their nodes in Zone I and the other half in Zone II. In addition, long trajectories also show a bigger popularity of Zone I over Zone II.

\subsection{Popularity of Each Node}
\label{sec_pop_of_loc}

In Fig.~\ref{fig_traj_cnt_and_popular_traj}, besides the popular trajectories, two sets of statistics are also depicted to show information about the popularity of each node. First, the ratio of trajectories passing by each node over all nodes (depicted as blue bars in the figure) shows the overall popularity of each node. Ratios of all ten nodes sum up to one. Among all the ten nodes, B, F, and C had the largest share of trajectories among all, which means the areas around them attracted the most visitors of the entire sample. Next, pie charts in the same figure show the ratio of single-node trajectories among all passing-by trajectories for each node, which reflects the attractiveness of each node. From the numbers, trajectories that passed by J, B, and F are more likely to be single-node trajectories, compared with other nodes. This indicates that these three nodes attracted visitors who were mainly visiting the surrounding areas instead of the event. In contrast, most of the visitors who visited nodes H, E, I, and D also visited other nodes.

\begin{figure}[!t]
\centering
\includegraphics[width=.95\linewidth]{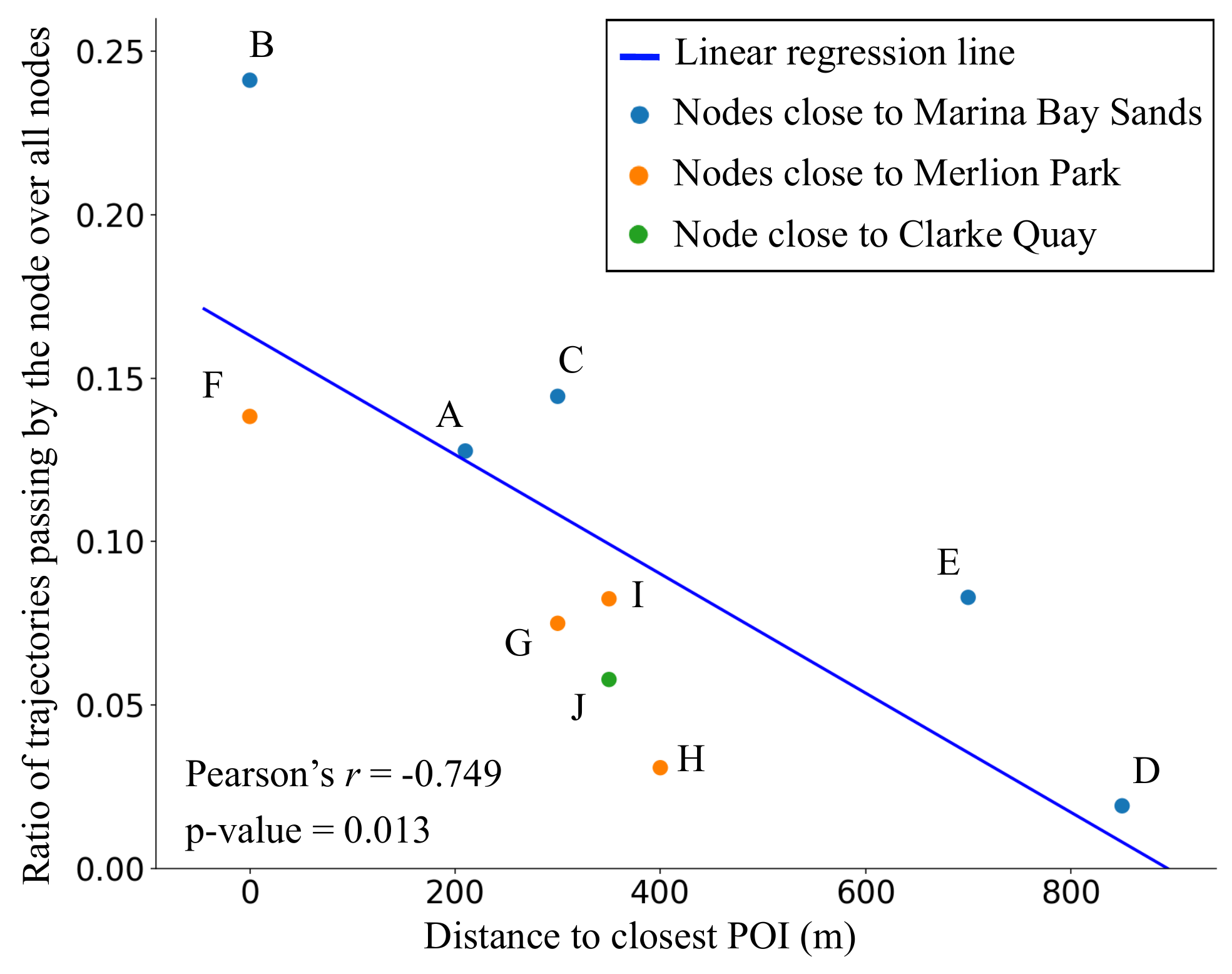}
\caption{Correlation between the ratio of trajectories passing by a node with its distance to closest place of interest (POI).}
\label{fig_traj_cnt_cause}
\end{figure}

Combining the above observations with the map of the event, it is found that the most popular or attractive parts of the event, namely areas around B, F, and J, are all close to a famous place of interest (POI) in Singapore, respectively. Node B is in front of Marina Bay Sands, node F is next to the Merlion statue in Merlion Park, and node J is nearby Clarke Quay. A hypothesis is thus raised that these POIs have a great impact on the popularity of each node. To quantitatively show this impact, analysis of the correlation between the ratio of trajectories passing by each node and its distance to the closest POI is conducted, and results are shown in Fig.~\ref{fig_traj_cnt_cause}. The linear regression line is obtained by fitting a linear model between the two variables with the method of least squares. From the line, there is a clear descending trend of the ratio of passing-by trajectories with the increase of the distance to the closest POI. Pearson correlation coefficient between the two variables and corresponding $p$-value are computed and shown in the figure, which verifies a significant negative impact of the distance to POI on the ratio of passing-by trajectories. From the scatters in the figure, one also can find that nodes nearby Marina Bay Sands are able to attract more visitors than those close to Merlion Park and Clarke Quay.

\subsection{Interconnections Between Nodes}
\label{sec_inter_nodes}

To investigate the interconnections between different nodes (i.e. how people transit from one node to the other), transition probability matrix $\boldsymbol{T}$ between the ten sensing nodes is computed from all the trajectories as:
\begin{equation}
\boldsymbol{T}(i,j) = \frac{\boldsymbol{N}(i,j)}{\sum_{k\in[1,10],k\not=i} \boldsymbol{N}(i,k)}
\end{equation}
where $\boldsymbol{T}(i,j)$ stands for the probability for a visitor to go from node $i$ to node $j$, $\boldsymbol{N}(i,j)$ stands for the times that a visitor went from node $i$ to node $j$ calculated from the trajectories, $i \not= j$, and thus $\sum_{k\in[1,10],k\not=i} \boldsymbol{T}(i,k) = 1$. To fill the diagonal of the matrix, when $i = j$, $\boldsymbol{T}(i,j)$ is set as one.

Hierarchical agglomerative clustering (HAC) is applied to automatically describe the interconnections between the nodes based on the transition probability matrix. HAC is a clustering algorithm that is designed to organize given data points into a hierarchical tree-like structure. At the first iteration of the algorithm, the input data points are paired based on the given distance measure, such as Euclidean distance, to form up small clusters. Then, at each of the following iterations, the clusters obtained in the previous iteration are further paired to form higher-level clusters based on given linkage criterion, which is a function of the pairwise distances between points inside the two clusters. Common linkage criteria include single, complete, average, weighted, and Ward linkage~\cite{mullner2011modern}. At the last iteration, all data points are finally connected into one cluster. During the algorithm, the leaf nodes of the organized tree (i.e. data points) are ordered such that distance between successive leaves is minimal, which helps the interpretation of the tree~\cite{bar2001fast}. From the obtained tree, one can observe the interconnections between data points and extract clusters of points at any desired level.

Here, the ten sensing nodes are clustered using HAC with the transition probability matrix $\boldsymbol{T}$ as the input feature matrix, and thus each row of the matrix, $\boldsymbol{T}(i,:)$, is the feature vector of the corresponding node $i$. Euclidean distance is adopted to measure the feature similarity between each pair of nodes, which represents the interconnection degree between the two nodes. The smaller the distance is, the stronger the interconnection is.
As for choosing the most suitable linkage criterion, the object at each iteration of the HAC is to pair two small clusters of nodes that are most interconnected with each other. To better describe the interconnection degree between two small clusters of nodes, the average of the interconnection degrees between all pairs of nodes from the two small clusters is selected.
In other words, the average linkage is chosen to measure the feature similarity (or the interconnection degree) between small clusters of nodes at each HAC iteration, which is computed as:
\begin{equation}
L(C_1,C_2) = \frac{1}{|C_1| \cdot |C_2|} \sum_{\boldsymbol{a} \in C_1} \sum_{\boldsymbol{b} \in C_2} \sqrt{\sum_{j} (\boldsymbol{a}_j - \boldsymbol{b}_j)^2}
\end{equation}
where $L(C_1,C_2)$ stands for the linkage criterion between two clusters of nodes $C_1$ and $C_2$, $|\cdot|$ is the cardinality of the cluster, $\boldsymbol{a}$ and $\boldsymbol{b}$ represents the nodes belonging to $C_1$ and $C_2$ respectively, and $j$ stands for the index of feature of each node (i.e. $j$th value of each row in $\boldsymbol{T}$).

\begin{figure}[!t]
\centering
\includegraphics[width=.95\linewidth]{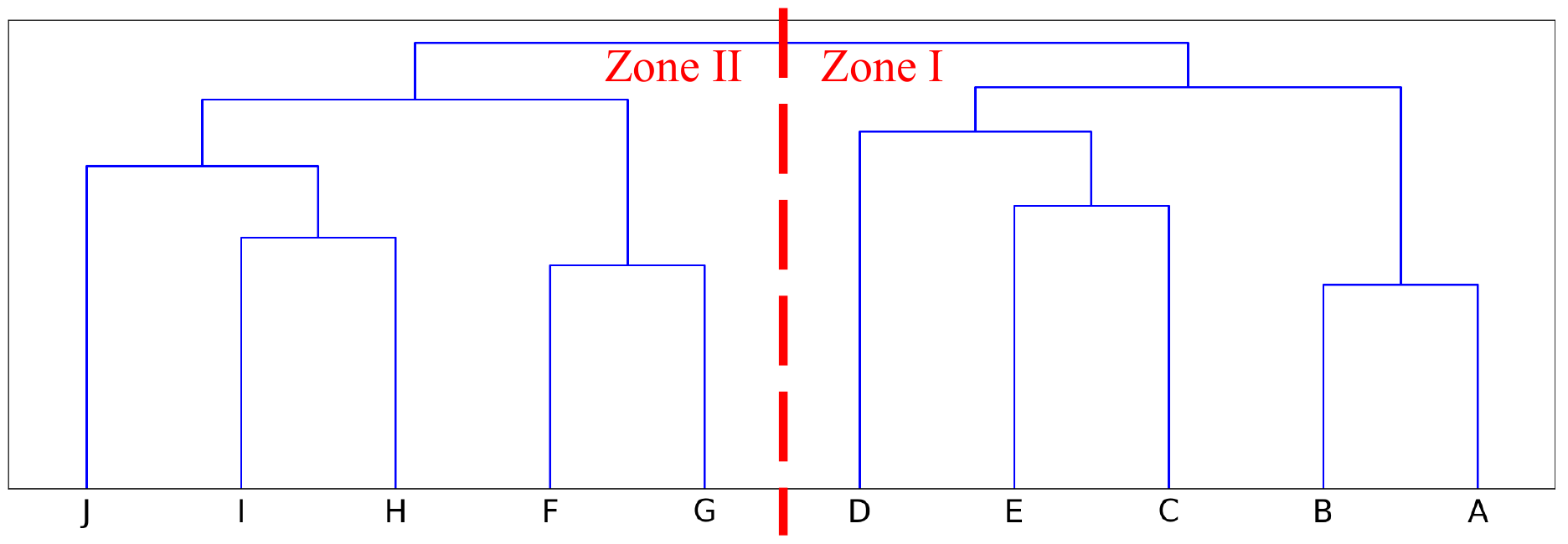}
\caption{Organized tree of nodes obtained by hierachical agglomerative clustering.}
\label{fig_hierarchical_clusters}
\end{figure}

The extracted organized tree of all the sensing nodes is depicted in the dendrogram in Fig.~\ref{fig_hierarchical_clusters}. From the figure, A-B, C-E, F-G, and H-I are the four pairs of nodes with the largest similarity, which is understandable as they are geographically close to each other. Subsequently, they are gradually connected with node D, node J and each other. Before the last combination, nodes from A to E form up one cluster, which corresponds to Zone I in Fig.~\ref{fig_traj_cnt_and_popular_traj}, and nodes from F to J form up another cluster, which corresponds to Zone II in Fig.~\ref{fig_traj_cnt_and_popular_traj}. This result thus mathematically reflects the previous observation on the separation of the entire event area (i.e. Zone I and Zone II) in Section~\ref{sec_com_traj}.

\begin{figure}[!t]
\centering
\includegraphics[width=1.0\linewidth]{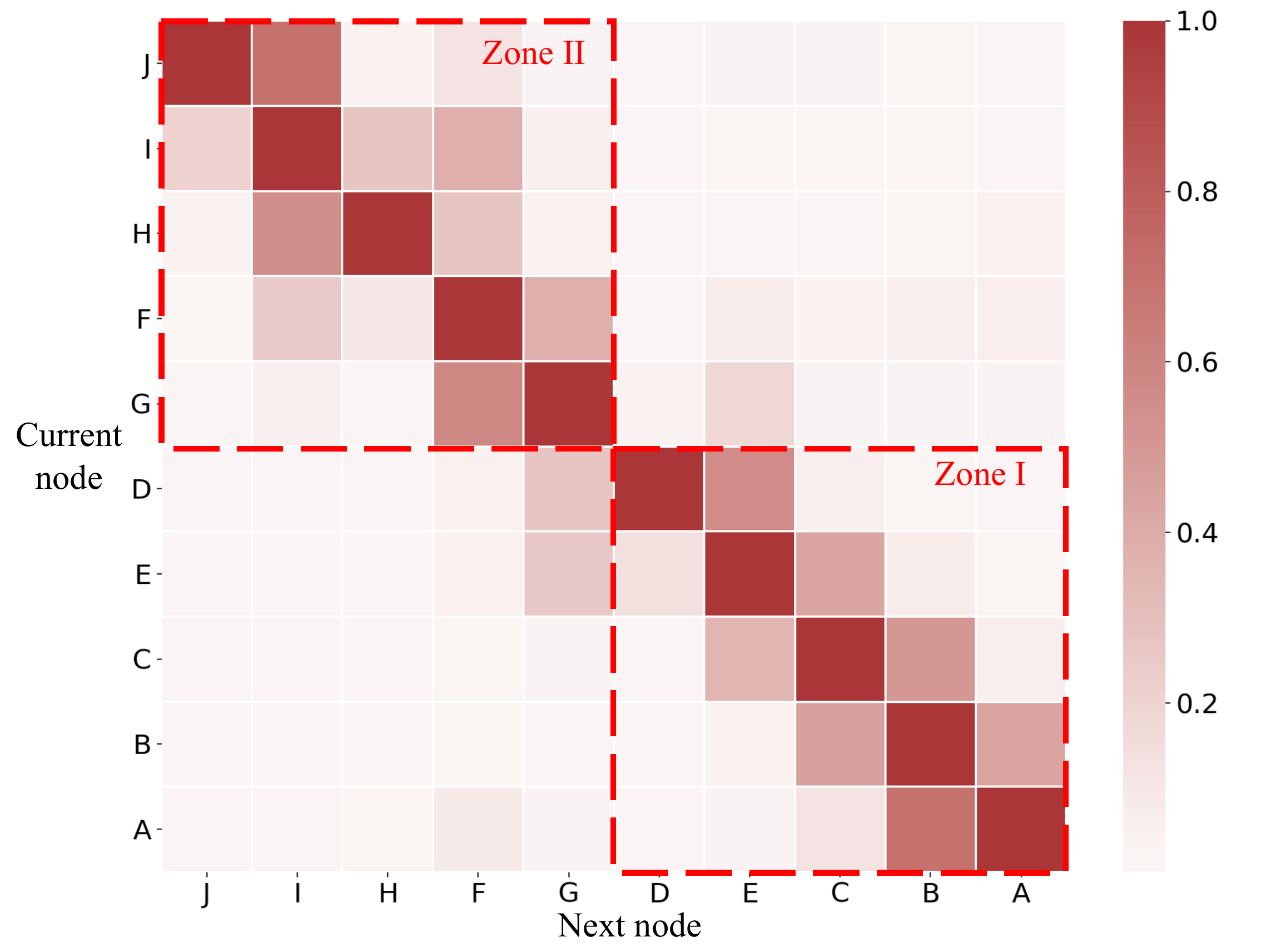}
\caption{Visualization of the transition probability matrix $\boldsymbol{T}$ between the ten sensing nodes. Sequence of nodes is organized based on the tree obtained by hierachical agglomerative clustering.}
\label{fig_transition_matrix}
\end{figure}

\begin{figure}[!t]
\centering
\includegraphics[width=1.0\linewidth]{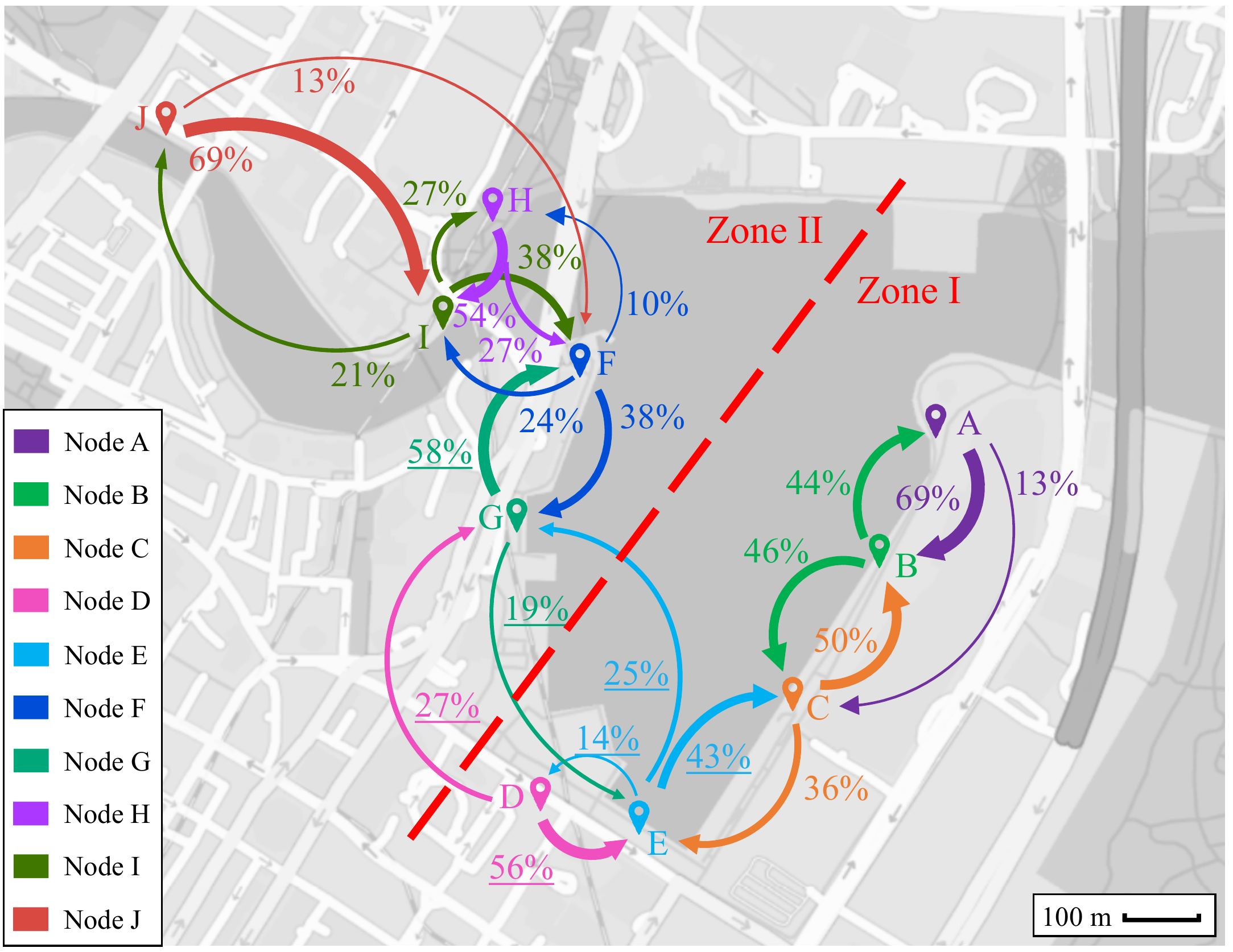}
\caption{Transition probability matrix $\boldsymbol{T}$ visualized by links of nodes. The width of each link indicates the value of $\boldsymbol{T}(i,j)$, and only links with $\boldsymbol{T}(i,j)$ larger than 10\% are plotted. Each color represents one node, and percentages with the same color are probabilities of going to other nodes from the node.}
\label{fig_transition_on_map}
\end{figure}

Based on the above tree of nodes, the transition probability matrix $\boldsymbol{T}$ is visualized by a heatmap in Fig.~\ref{fig_transition_matrix} (with the sequence of nodes reorganized) and by links between nodes in Fig.~\ref{fig_transition_on_map}. Zone I and Zone II are marked by red dash lines in the two figures. In Fig.~\ref{fig_transition_matrix}, the color of a block indicates the value of corresponding $\boldsymbol{T}(i,j)$. In Fig.~\ref{fig_transition_on_map}, the width of link between nodes indicates the value of $\boldsymbol{T}(i,j)$, and only links with $\boldsymbol{T}(i,j)$ larger than 10\% are plotted. From the two figures, it is clear that people were most likely to transit to another node following the geographical adjacency.
In other words, starting from one node, the majority of visitors would go to the nodes right next to the current node instead of jumping to further nodes. Moreover, from Fig.~\ref{fig_transition_on_map}, one can see that the connections between Zone I and Zone II mainly happened in the links of E-G and D-G. At nodes D, E, and G, the probability of crossing into the other zone for the visitors is around 27\%, 25\%, and 19\% respectively. As shown by these numbers, when people were at the edge of one particular zone (i.e. node D, E, and G), they were more likely to stay in the current zone rather than entering the other zone. This could be caused by the long distances between node G and node D, E as well as the possibility that there is not a very attractive art installation in between.

Although the above results on interconnections between nodes by HAC seem intuitive because the sensing nodes are well organized by the surrounding geography (i.e. the bay), its effectiveness in obtaining such interconnections is verified. For passive WiFi sensor networks deployed with more nodes and without clear geographical boundaries, HAC can be extremely useful.
For example, in~\cite{kalogianni2015passive} and~\cite{acer2016capturing}, 30 and 20 WiFi monitors were deployed in a large-scale indoor exhibition event and a university campus respectively. In these two experiment settings, there is no geographical boundary that restricts people's walking routes like the bay area in this study. In other words, people can walk from one location to the other freely, not having to follow a certain sequence of locations. As a result, the interconnections between monitored locations are no longer intuitive. In such situations, HAC can be applied to obtain the hidden interconnections between locations automatically.

\section{Temporal Patterns}
\label{sec_temporal}

Two types of temporal patterns of crowd behaviors during the event are extracted in this section, namely daily patterns of overall visitor count and patterns of visitor count versus time at each node. We use mobile device count to obtain these patterns because it is reasonable to assume that the trend of device count is consistent with the trend of visitor count. As fine temporal granularity is required to extract temporal patterns and no movement information is needed in this stage, \textit{Dataset A} is used for the analysis, including probe request records with both global and local MAC addresses. Since only relative trends of device counts are of interest, using all the probe request records is acceptable, assuming that the mapping rate from MAC addresses to mobile devices is consistent under the studied temporal granularity. The temporal granularity is set as 15 minutes in the following analysis.

\subsection{Daily Patterns of Overall Device Count}

The entire data collection lasted for 27 days and covered all types of days, namely workdays, weekends, public holidays (PHs), and festivals but not PH. It is expected that on different types of days, the count of visitors (devices) should have different magnitudes. Therefore, to extract daily patterns of the overall device count, magnitude is the main basis. But, to differentiate between days with similar magnitudes, shapes of the curves of device count versus time need to be considered as well. As a result, $k$-means clustering is applied to the daily curve of overall device count versus time. In other words, each day is considered as one data point and device counts at different timestamps on the day form up the feature vector of the data point. To extract such a feature vector, the overall count of detected unique MAC addresses is computed for every time interval. Subsequently, the values of the count of unique MAC addresses are normalized over all days to the range of $[0,1]$ by min-max feature scaling. In other words, the count of unique MAC addresses at each time interval for each day is first subtracted by the global minimum count (minimum over all time intervals of all days) and then divided by the difference between the global maximum and global minimum.

$k$-means clustering is able to group input data points (i.e. the 27 days) into $k$ clusters in an expectation-maximization (EM) schema. After proper initialization, at each iteration, the algorithm assigns a label to each day based on present cluster $k$ centroids and then updates the centroids by taking the arithmetic mean of feature vectors of days in each corresponding cluster. After the algorithm converges, $k$ clusters of days are obtained.
To select the proper value of $k$, the usual practice is to compare the values of a certain criterion under different values of $k$ and select the $k$ with the best criterion value. Among the available criteria, distance-based criteria, such as the within-cluster sum of squared distance and Silhouette value, are easy to compute, easy to understand and suitable for small-size datasets. Moreover, if using the within-cluster sum of squared distance as the criteria, when k increases, the criterion keeps increasing. One then needs to use the Elbow method to select the $k$ which can be very ambiguous and subjective when the data are not well separable. In contrast, Silhouette value combines the within-cluster distances and cross-cluster distances and thus avoids the use of the Elbow method~\cite{rousseeuw1987silhouettes}. Just by comparing the mean Silhouette values over all data points, one can select the proper value of $k$.
In this case, Silhouette value of each clustered day measures how close the particular day is to all other days in the same cluster and how far the day is from days in other clusters. For a tested $k$ value, the Silhouette value of a given day $i$, $s(i)$, is computed with following equations:
\begin{equation}
\begin{gathered}
s(i) = \left\{ \begin{array}{ll}
            \frac{d_{ex}(i)-d_{in}(i)}{\max \lbrace d_{in}(i), d_{ex} (i) \rbrace} & \mbox{if $|C_i|>1$}; \\
            0 & \mbox{if $|C_i|=1$}. \end{array} \right.
\\
d_{in}(i) = \frac{1}{|C_i|-1} \sum_{j \in C_i, i \not =j}d(i,j)
\\
d_{ex}(i) = \min_{k \not = i} \frac{1}{|C_k|} \sum_{j \in C_k}d(i,j)
\\
d(i,j) = \sqrt{\sum_{t} \lbrack n_i(t) - n_j(t) \rbrack ^2}
\end{gathered}
\end{equation}
where $d_{in}$ is the mean distance between day $i$ and other days in the same cluster, $d_{ex}$ is the mean distance between day $i$ to days in the closest neighbour cluster, $C_i$ is the cluster where day $i$ is in, $C_k$ is a neighbour cluster to day $i$, $|\cdot|$ is the cardinality of the cluster, $d(i,j)$ is the distance measure between day $i$ and day $j$, and $n_i(t)$ stands for normalized device count at time $t$ of day $i$. Moreover, since initialization of $k$-means can also have a big impact on the clustering results, 50 trials with the same $k$ value is conducted and trial with best mean Silhouette value is reserved.

\begin{figure}[!t]
\centering
\includegraphics[width=.9\linewidth]{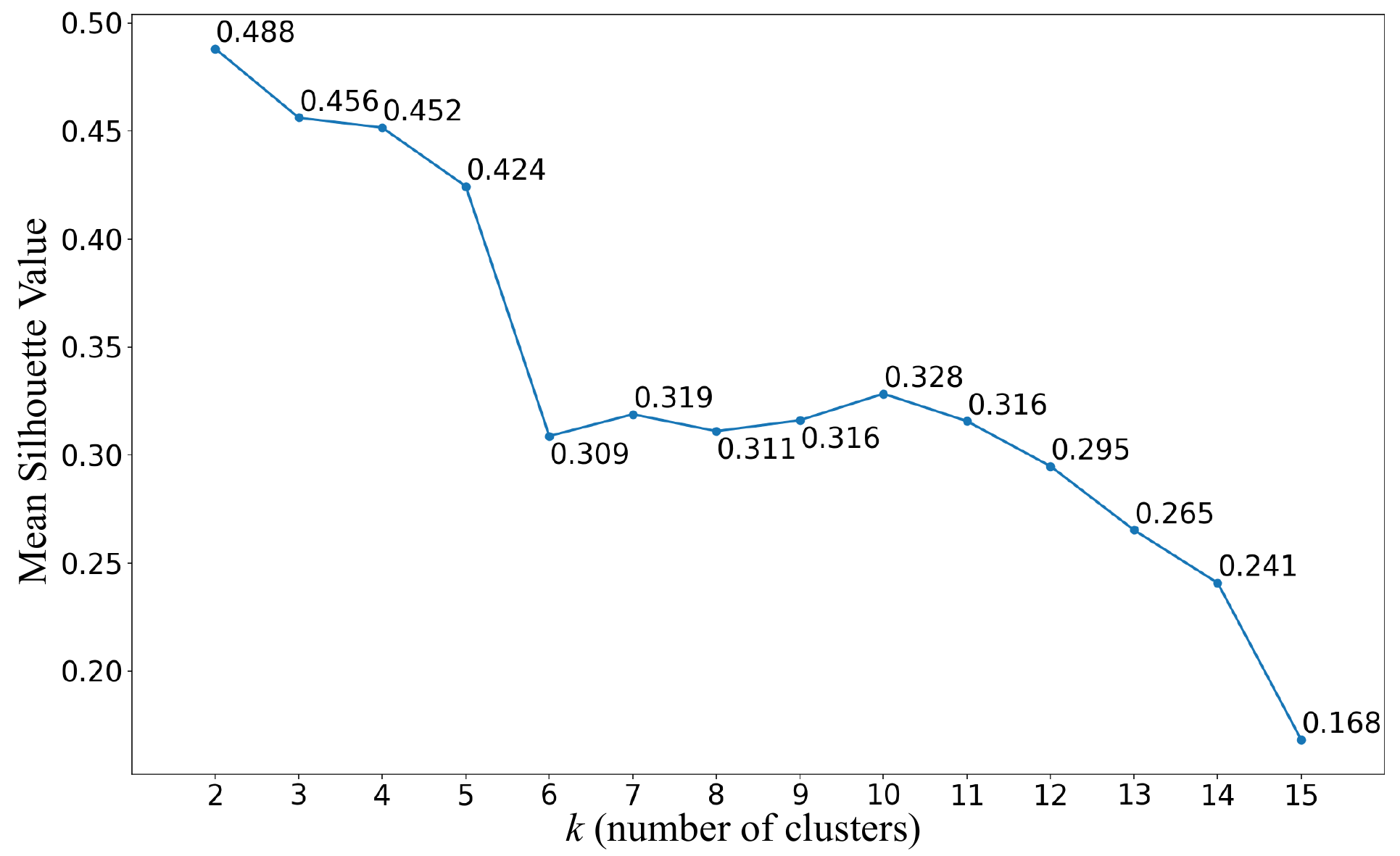}
\caption{Mean Silhouette value over all days, generated by $k$-means clustering with different $k$ values.}
\label{fig_silhouette_vs_k}
\end{figure}

In Fig.~\ref{fig_silhouette_vs_k}, the mean Silhouette value over all days obtained with different $k$ values are shown. From the figure, the separation of clusters is maximized when $k$ equals 2 and remains at a high level when $k$ equals 3, 4 and 5. The mean Silhouette values corresponding to three and four clusters are very close. After visualizing and interpreting these four sets of results, $k=4$ is selected to balance the number of discovered patterns (clusters) and the similarity between data points within the same cluster.

\begin{figure}[!t]
\centering
\subfloat[]{\includegraphics[width=.8\linewidth]{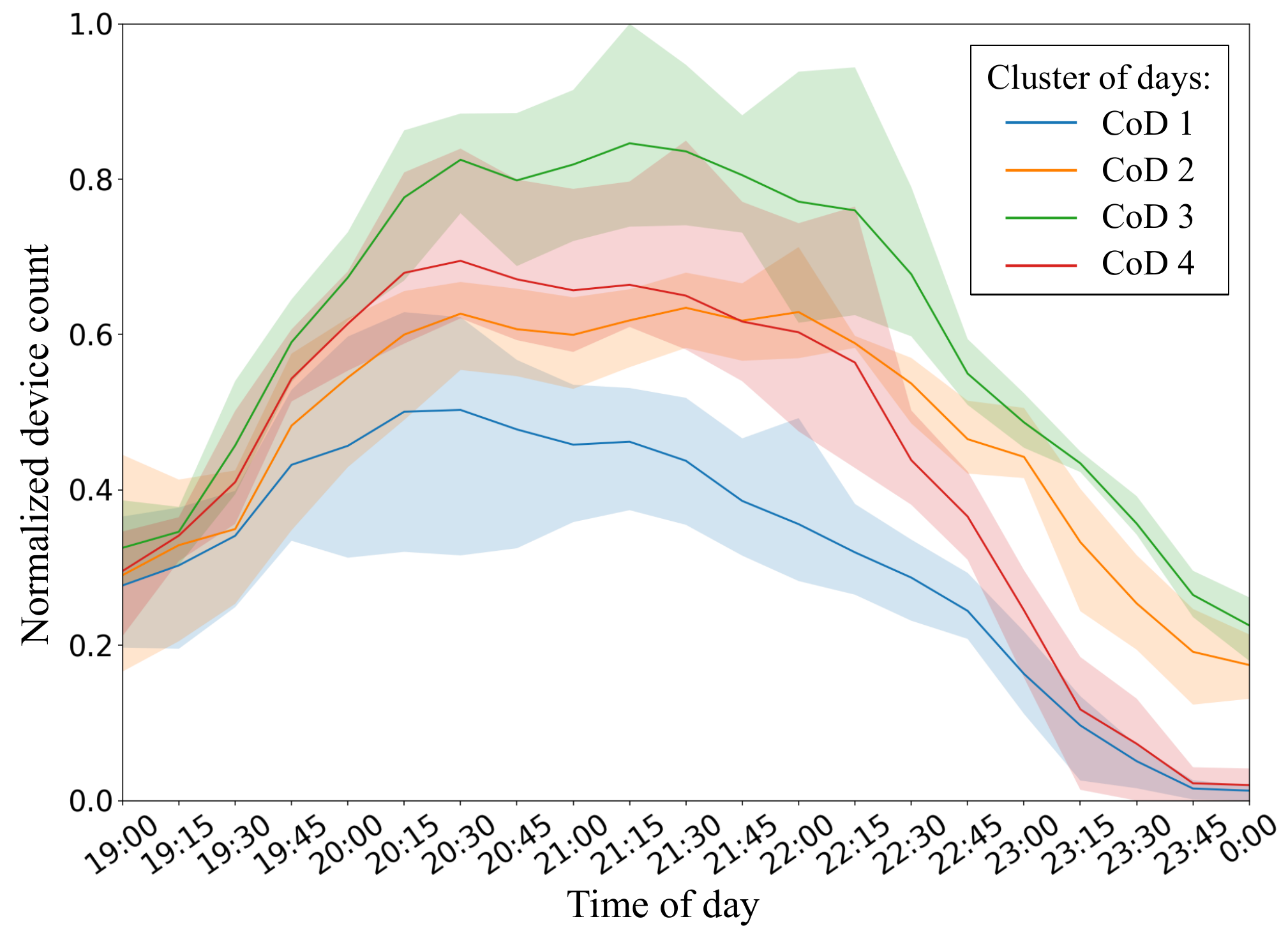}%
\label{fig_device_cnt_by_clusters_of_days}}
\hfil
\subfloat[]{\includegraphics[width=.8\linewidth]{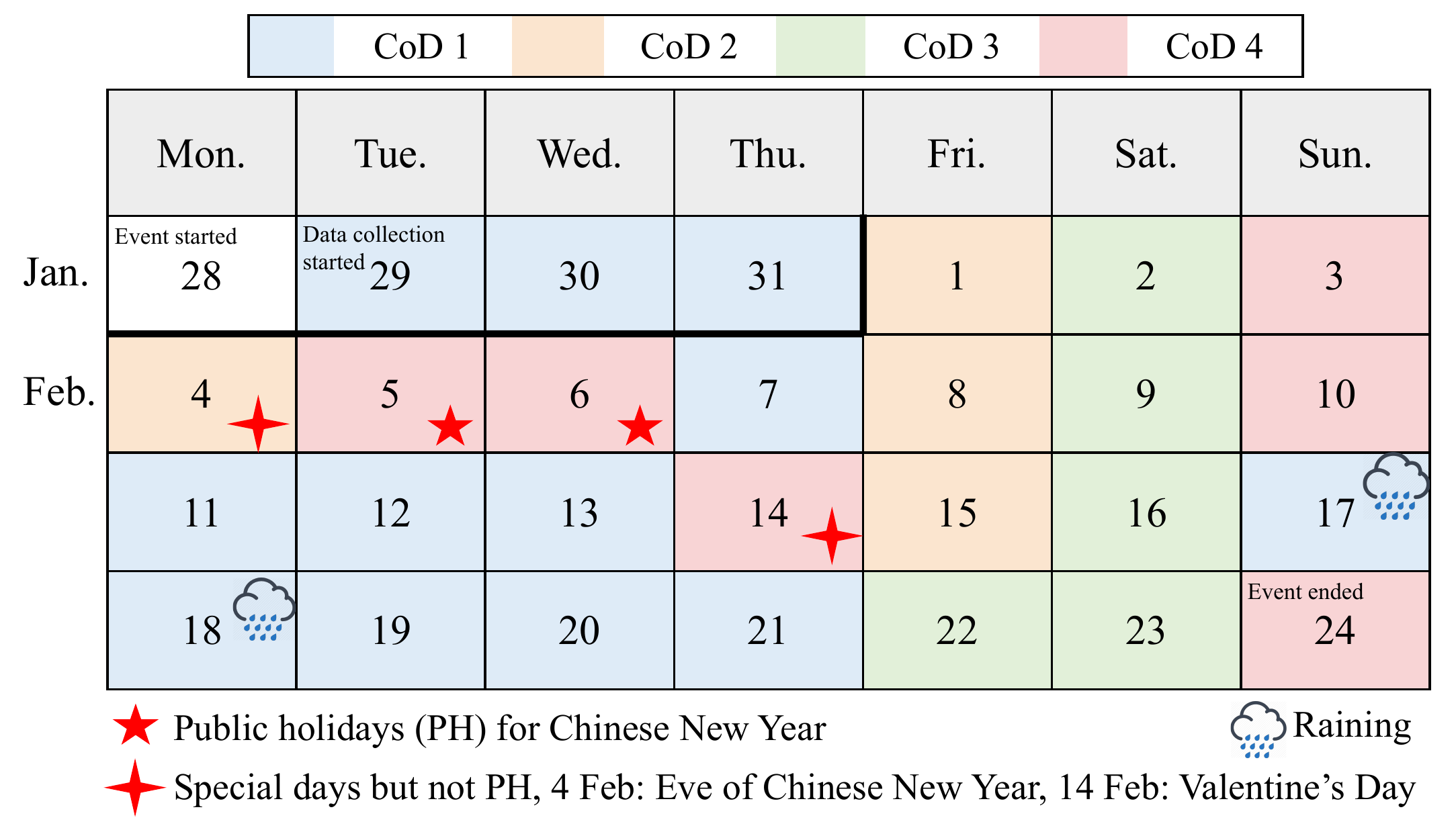}%
\label{fig_calendar}}
\caption{Four clusters of days (CoD 1-4) obtained from $k$-means clustering and their temporal patterns of overall device count: (a) curve of normalized device count versus time of day of each cluster, solid line in the middle represents the mean value over the cluster, and shading area represents the range between the maximum and minimum values within each cluster, (b) cluster label of each day shown in a calendar.}
\label{fig_clustering_by_days}
\end{figure}

Four clusters of days are, therefore, produced as the final result of $k$-means clustering. In Fig.~\ref{fig_device_cnt_by_clusters_of_days}, the daily pattern of normalized device count of each cluster is plotted in different colors. In Fig.~\ref{fig_calendar}, the four clusters of days (CoD 1-4) are visualized in a calendar with each day colored based on its cluster label. From the two figures, the obtained four clusters mainly correspond to four types of days. For days in CoD 1 (mainly including Mondays to Thursdays), they have the lowest magnitude of device count overall and their device count starts to decline early of the day. For days in CoD 2 (mainly including Fridays), they have the median magnitude of device count overall and their device count decreased slowly at late night. For days in CoD 3 (mainly including Saturdays), they have the highest magnitude of device count overall and their device count decreased slowly at late night. As for days in CoD 4 (mainly including PHs and Sundays), they have the median magnitude of device count overall, same as CoD 2, but their device count decreased sharply after 10 pm.

Besides the above overall patterns, the clustering also captures interesting knowledge about individual days as shown by Fig.~\ref{fig_calendar}. For the two PHs of Chinese New Year (CNY), 5th and 6th February, they had daily device count patterns similar to Sundays (CoD 4). For 4th and 14th February, which is the CNY eve and the Valentine's Day (not public holidays), they are clustered into CoD 2 and 4, respectively. Because the next day of 4th was a PH, it showed a daily pattern similar to Fridays. As for 14th, as its next day was a working day, it showed a daily pattern similar to Sundays. Moreover, two days that are abnormal to their day of the week are detected, namely 17th and 22nd February. On 17th, it was raining, so the device count is expected to be lower than normal Sundays, as a result, it is clustered with other ordinary workdays. As for 22nd, it showed a daily pattern similar to Saturdays rather than Fridays, which means a higher magnitude of device count than usual Fridays. This can be explained by the fact that the i Light Singapore event ended on 24th February, which boosted the count of visitors on 22nd. For 24th itself, since the next day was Monday, although the magnitude of device count could be higher than usual due to the ending of the event, it is still clustered into CoD 4 because it showed a sharp decline after 10 pm.

\subsection{Patterns of Device Count at Each Node}

Besides the daily patterns of overall device (visitor) count, how people visited each node at different time intervals of the day is also an interesting question to investigate. Since four clusters of days with similar patterns have been extracted from the above analysis, the investigation of device count versus time of day at each node should be conducted for each cluster of days separately. However, for CoD 2-4, they have too few members, and the stochastic nature of crowd behaviors would have a large impact on the obtained patterns. Therefore, only CoD 1 is selected for this analysis, since it has the largest number of members (12) and most of them are ordinary workdays.
For all days within CoD 1, their values of device count during one particular time interval of the day at each node are averaged and used to represent the entire cluster. By doing so, the stochastic nature of crowd behaviors across days is relieved.

\begin{figure}[!t]
\centering
\includegraphics[width=.95\linewidth]{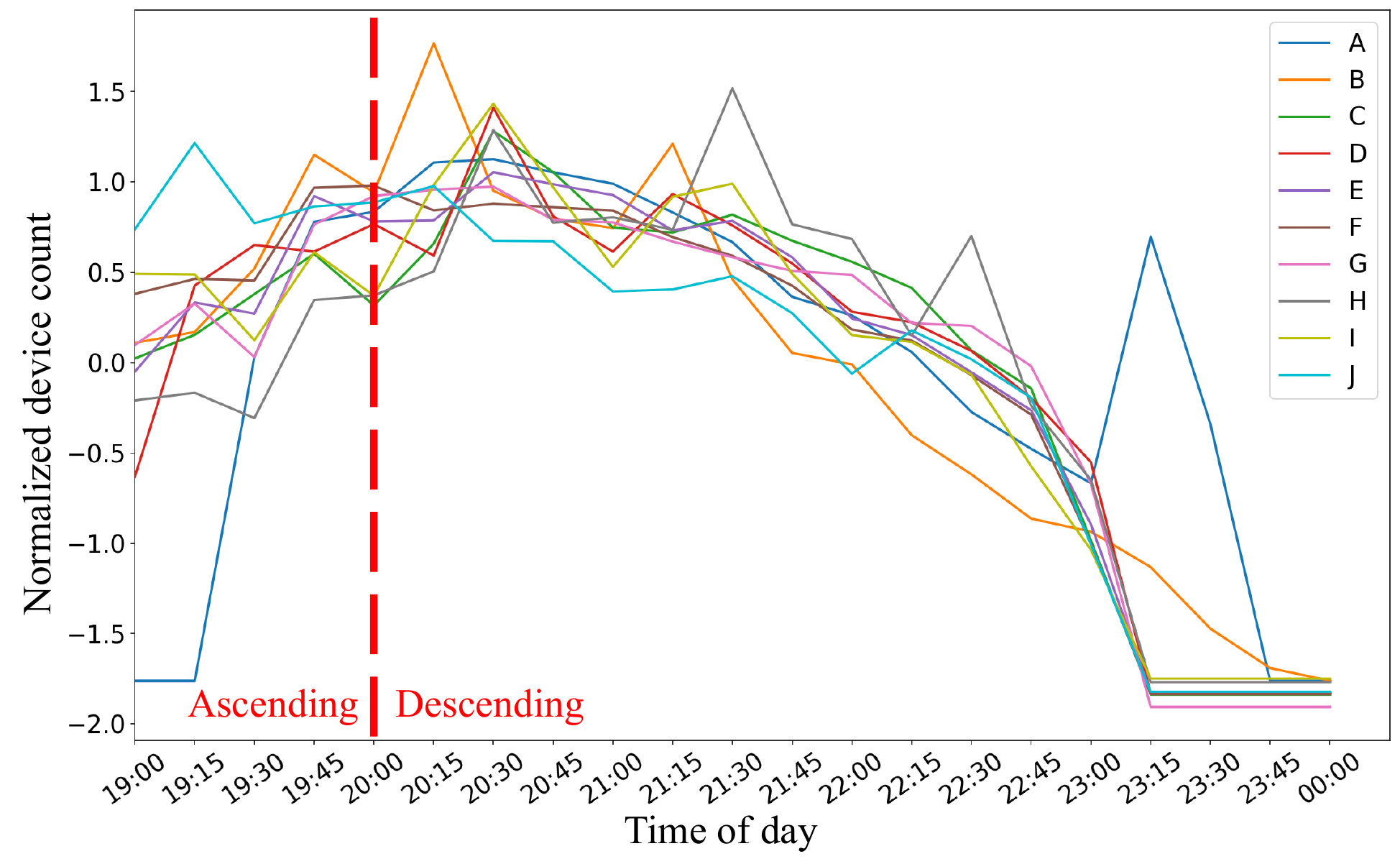}
\caption{Normalized device count versus time at each node averaged over days in CoD 1.}
\label{fig_device_cnt_perts_all_nodes}
\end{figure}

Because each node has different popularity as shown in Section~\ref{sec_pop_of_loc}, the magnitude of device count at each node is very different from each other. Therefore, shapes of device count curves (vs. time) of the ten nodes are the focus of the analysis instead of their magnitudes. To extract patterns from these curves based on their shapes, $k$-shape clustering is adopted, which is an effective shape-based clustering method for time-series data proposed by John Paparrizos and Luis Gravano~\cite{paparrizos2015k}. It is useful to differentiate curves based on their partial shapes when the curves have similar overall trends and possible phase shifts.

In Fig.~\ref{fig_device_cnt_perts_all_nodes}, the curve of average device count versus time at each node over days in CoD 1 is shown. Each curve is normalized by itself with z-normalization (i.e. subtracting its mean value and dividing by its standard deviation), which is required by the $k$-shape clustering to make sure the same scale for all curves. From the figure, one can find that the curves of normalized device counts at the ten nodes are close to each other but with different partial shapes (e.g. local peaks), which are the targets to capture. Moreover, the overall trends of all curves can be roughly divided into two parts: quick ascending from 7 pm to 8 pm and gradual descending from 8 pm to 12 am. Since there are only ten nodes to cluster, to avoid generating too many clusters with a single member, the gradual descending parts during 8 pm to 12 am are extracted and clustered to capture the patterns of how visitor count changed at different nodes.
The values of these parts (i.e. 8 pm - 12 am) of the curves are normalized to z-scores again before conducting the clustering.

$k$-shape clustering, which is designed to cluster time-series data based on their shapes, shares a similar iterative refinement (or optimization) procedure as the $k$-means clustering, but it adopts different distance measure and method of computing cluster centroids. For a given pair of time-series vectors with same length, $\boldsymbol{x}$ and $\boldsymbol{y}$, the similarity between them is measured by shape-based distance (SBD) computed as:
\begin{equation}
\begin{gathered}
SBD(\boldsymbol{x},\boldsymbol{y}) = 1 - \max_{s} \left(\frac{CC_s(\boldsymbol{x},\boldsymbol{y})}{\|\boldsymbol{x}\| \cdot \|\boldsymbol{y}\|} \right)
\\
CC_s(\boldsymbol{x},\boldsymbol{y}) = \left\{ \begin{array}{ll}
                                        \sum_{l=1}^{m-s} \boldsymbol{x}_{l+s} \cdot \boldsymbol{y}_l & \mbox{if $s \geq 0$}; \\
                                        CC_{-s}(\boldsymbol{y},\boldsymbol{x}) & \mbox{if $s < 0$}. \end{array} \right.
\end{gathered}
\end{equation}
where $CC_s(\boldsymbol{x},\boldsymbol{y})$ stands for the cross-correlation between $\boldsymbol{x}$ and $\boldsymbol{y}$ with a index shift $s$, $s \in \lbrack -m, m \rbrack$, $m$ is the length of each vector, and $\|\cdot\|$ stands for Euclidean norm of the given vector. From above equations, SBD measures the similarity between the most similar parts from $\boldsymbol{x}$ and $\boldsymbol{y}$. As for the computation of centroid of each cluster at each iteration of $k$-shape clustering, an optimization problem is solved to find the vector $\boldsymbol{\mu}^*$ which has the minimum sum of SBD to all vectors inside the corresponding cluster. Details of implementation can be found in~\cite{paparrizos2015k}.

\begin{figure}[!t]
\centering
\subfloat[]{\includegraphics[width=1.0\linewidth]{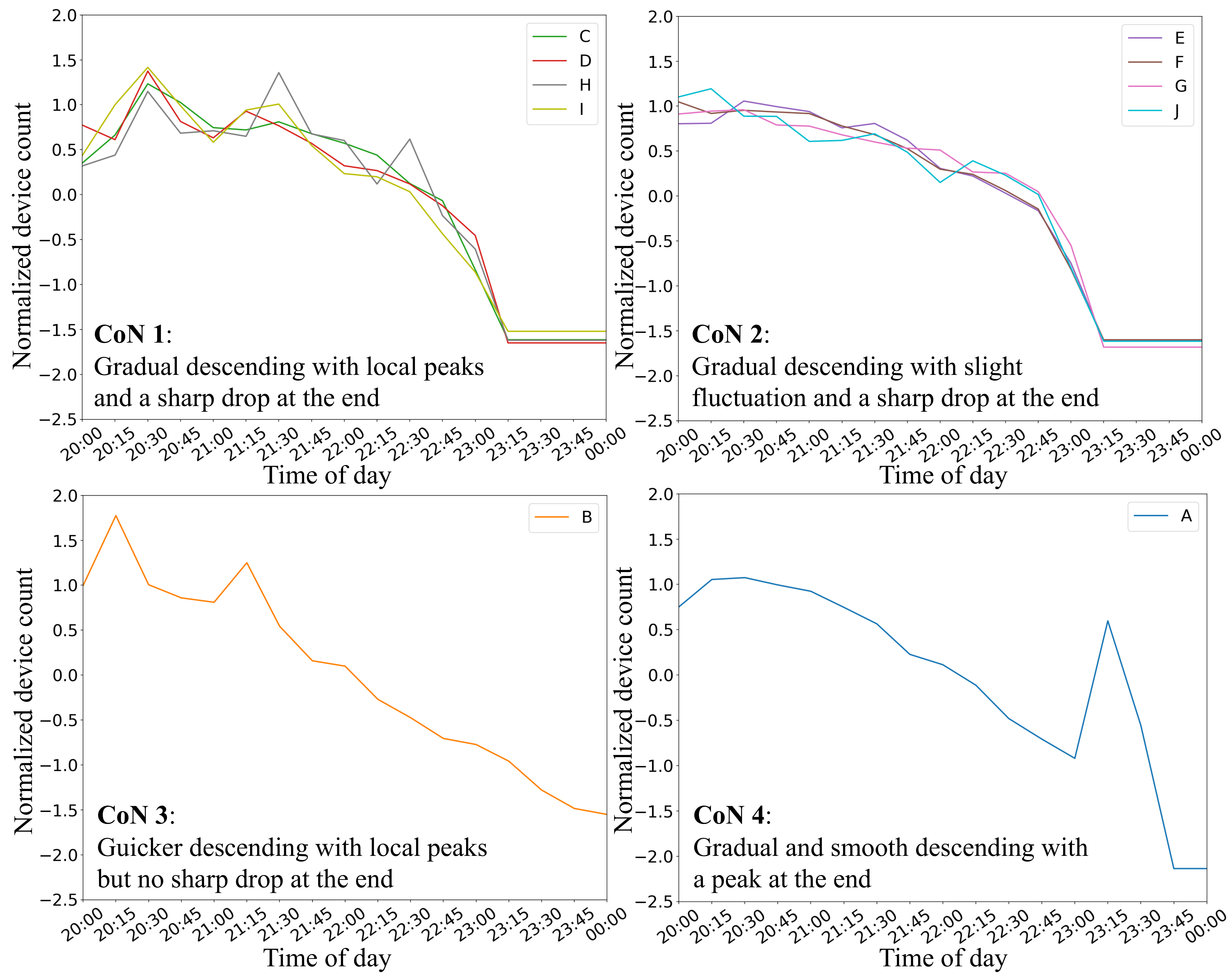}%
\label{fig_kshape_night}}
\hfil
\subfloat[]{\includegraphics[width=.7\linewidth]{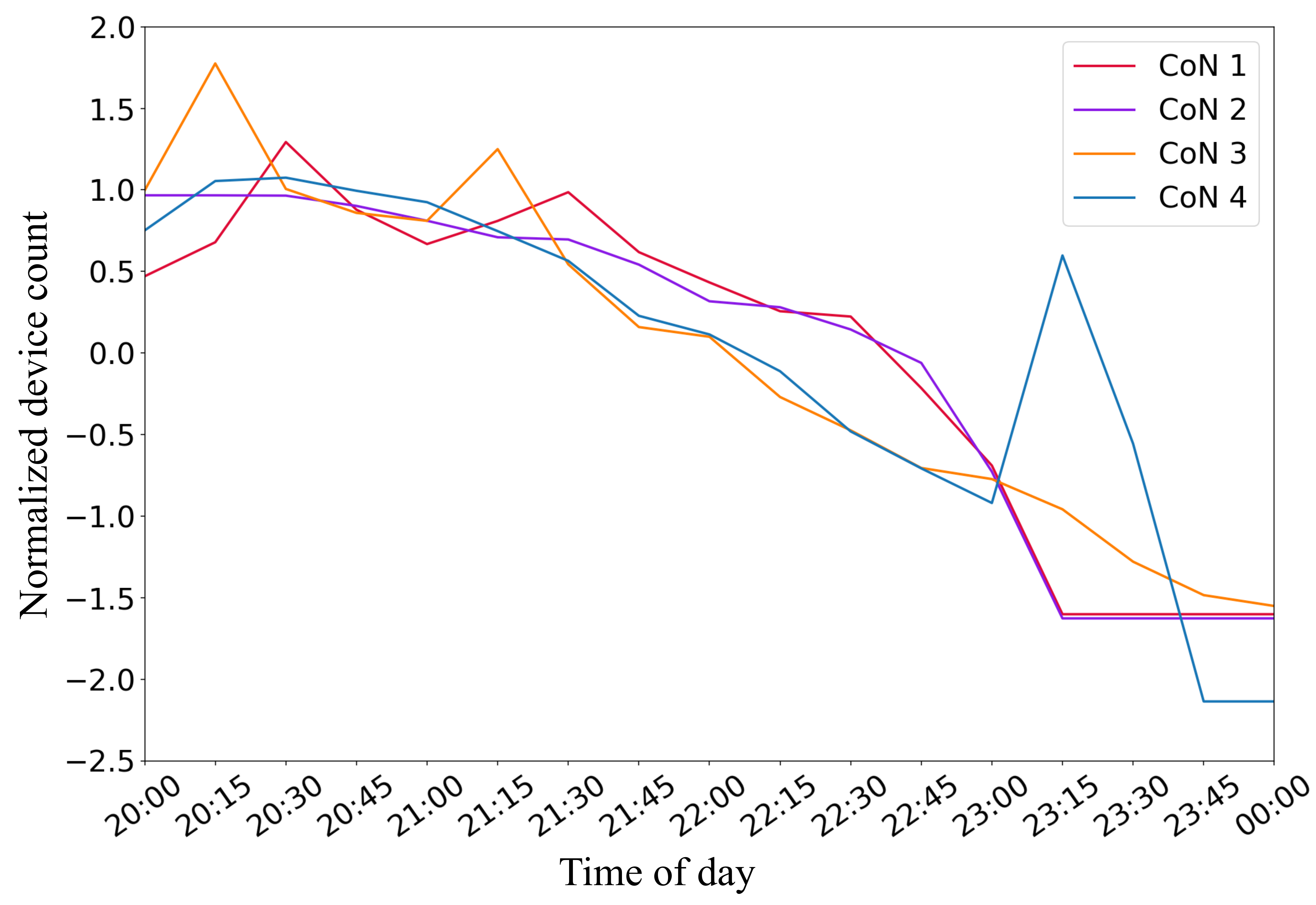}%
\label{fig_kshape_night_mean_values}}
\caption{Results of $k$-shape clustering: (a) four clusters of nodes (CoNs) based on their normalized device count curves from 8 pm to 12 am, (b) mean of normalized device count curves of nodes in each CoN.}
\label{fig_kshape_results}
\end{figure}

After applying $k$-shape clustering to the curves of normalized device count from 8 pm to 12 am of all the nodes, four clusters of nodes (CoNs) are obtained. Here, since only ten time-series vectors (corresponding to the ten nodes) are clustered and the differences of SBD values between them are not large, the Silhouette measure (with SBD as the distance measure) is not able to tell the performance differences between different $k$ values. Instead, manual inspection and selection are conducted, and $k=4$ is selected.

In Fig.~\ref{fig_kshape_night}, the normalized device count curves of members (nodes) of each CoN are shown. In Fig.~\ref{fig_kshape_night_mean_values}, the mean of normalized device count curves of nodes in each cluster is shown. From these two figures, one can see that the obtained four clusters represent four different patterns of visiter count versus time.
For nodes in CoN 1 and 2, their device count curves share the same overall trend: gradually descending with a sharp drop around 10:45 pm, which can be understood as most of the visitors started to leave the event area. The main difference between these two clusters is that curves in CoN 1 have large local peaks and those in CoN 2 are either smooth or with only small fluctuations.
As for CoN 3 and 4, the clustering algorithm captured two nodes with special patterns. The device count curve of node B has two large peaks around 8:15 pm and 9:15 pm, and besides this, it decreases with a relatively quick and steady speed. As for node A in CoN 4, its device count shows an unexpected peak at late night (11:15 pm), which could be caused by common routes people took to leave the event area.
Moreover, from the curves in Fig.~\ref{fig_kshape_night}, five nodes (B, C, D, H, I) show obvious local peaks around 8 pm, 9 pm, or 10 pm. This observation can be explained by the special light shows that were scheduled near node I at 8 pm, 9 pm, and 10 pm, and near node B at 8 pm and 9 pm.

\section{Spatiotemporal Patterns}
\label{sec_spatiotemporal}

In this section, the spatiotemporal patterns of crowd movement are investigated in two different angles: how the trajectory duration changes with trajectory length (number of nodes visited), and how crowd movement changes over time daily. To extract such patterns, \textit{Dataset B} is used.

\subsection{Change of Trajectory Duration over Length}

\begin{figure}[!t]
\centering
\includegraphics[width=0.9\linewidth]{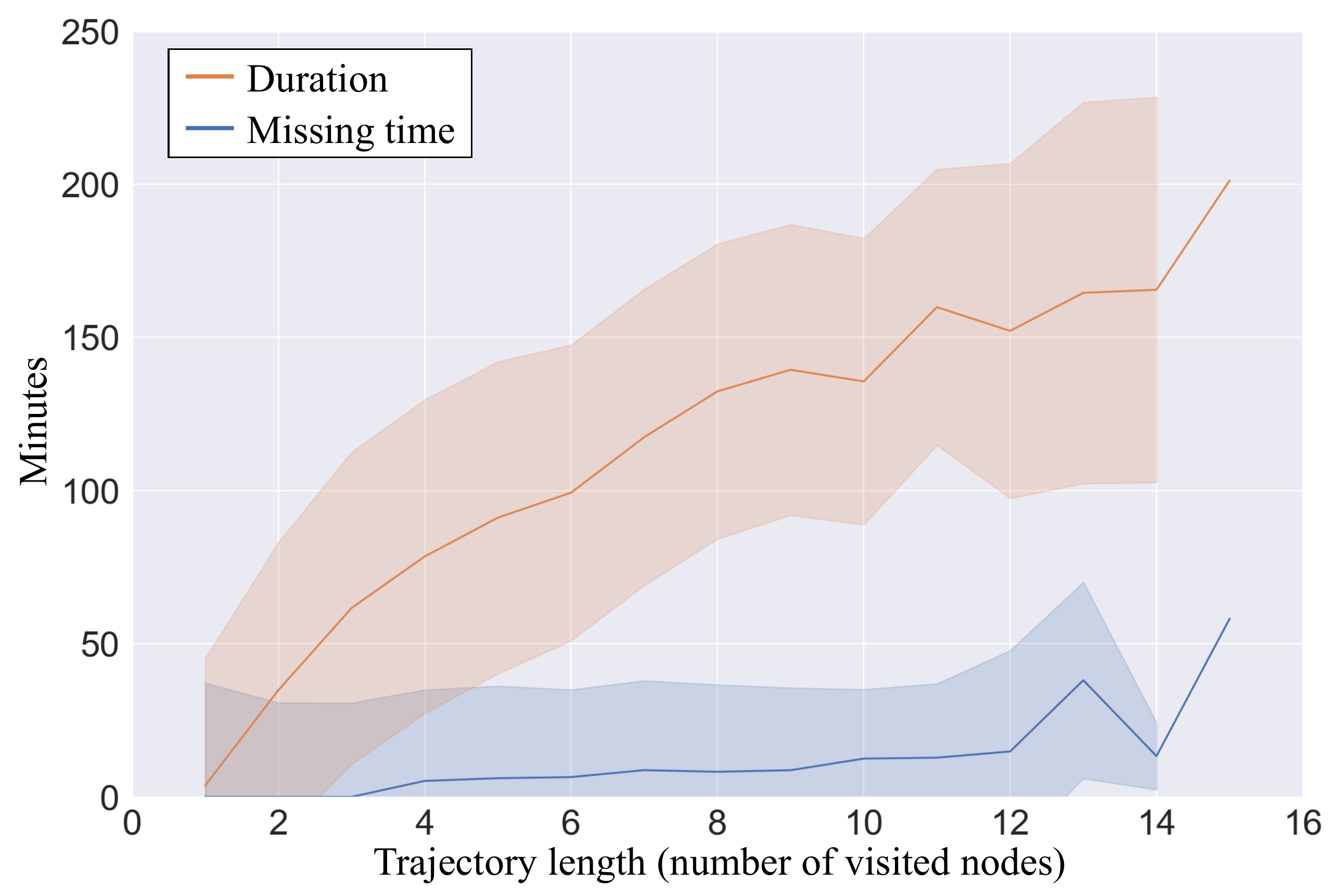}
\caption{Median (solid lines) and standard deviation (shading area) values of duration and missing time of trajectories with different number of visited nodes. When the trajectory length increases, sample size decreases exponentially. It just represents that missing time has larger variation than duration. When the length is 15, the sample size becomes one, and thus no standard deviation value is computed.}
\label{dist_of_duration_over_traj_length}
\end{figure}

Trajectory duration is defined as the total time the corresponding person spent inside the event area, which is computed as the span between the start time of the first node and the end time of the last node. To study the change of trajectory duration over trajectory length, typical values of duration of trajectories with different numbers of visited nodes are computed and visualized in Fig.~\ref{dist_of_duration_over_traj_length} (the orange part). From the figure, the median value of trajectory duration increases when the trajectory length increases, and the increasing trend slows down gradually. In addition, the standard deviation values show that, given the same number of visited nodes, there is a large variety of time spent in the event area by different people.

As introduced in Section~\ref{sec_data_preprocess}, sometimes, people left a node, remained undetected for a period, and finally showed up again at the same node. This period of being undetected is defined as the missing time in the data preprocessing stage. As supplementary information to the trajectory duration, typical values of total missing time of trajectories with different numbers of visited nodes are computed and visualized in Fig.~\ref{dist_of_duration_over_traj_length} (the blue part). From the figure, with the increasing trajectory length, the median of missing time increases slightly, while the standard deviation decreases. Moreover, when a trajectory is short, the portion of missing time over the duration is large, which means that the person is more likely to spend more time exploring the area near the visited nodes. In contrast, when the trajectory is longer, the portion of missing time over the duration becomes smaller, which represents that the person is more likely to stick with the trajectory during his/her visit to the event.

\subsection{Change of Crowd Movement over Time}

To investigate how crowd movement change over time daily, first, the daily data collection time (i.e. from 7 pm to 12 am) is divided into four periods: 7-8 pm, 8-9 pm, 9-10 pm, 10 pm - 12 am. Next, for each time period, the count of transitions from node $i$ to node $j$, $\boldsymbol{N}_t(i,j)$ is computed from all trajectories. Subsequently, for the link of node $i$ and $j$, ratio of each direction (from $i$ to $j$ and from $j$ to $i$) of the link is computed as:
\begin{equation}
\label{equa_direct_r}
R_t(i,j) = \frac{\boldsymbol{N}_t(i,j)}{\boldsymbol{N}_t(i,j)+\boldsymbol{N}_t(j,i)},
R_t(j,i) = \frac{\boldsymbol{N}_t(j,i)}{\boldsymbol{N}_t(i,j)+\boldsymbol{N}_t(j,i)}
\end{equation}
where $R_t(i,j)+R_t(j,i)=1$. From $R_t(i,j)$, one can determine which direction is the main direction for every link of two nodes. When $|R_t(i,j)-R_t(j,i)|<10\%$, it is defined as a mutual status for the link between $i$ and $j$, meaning no dominant link direction existing at time t. When $|R_t(i,j)-R_t(j,i)| \geq 10\%$, dominant link direction is extracted to represent the main trend of crowd movement at the particular link.

\begin{figure}[!t]
\centering
\includegraphics[width=1.0\linewidth]{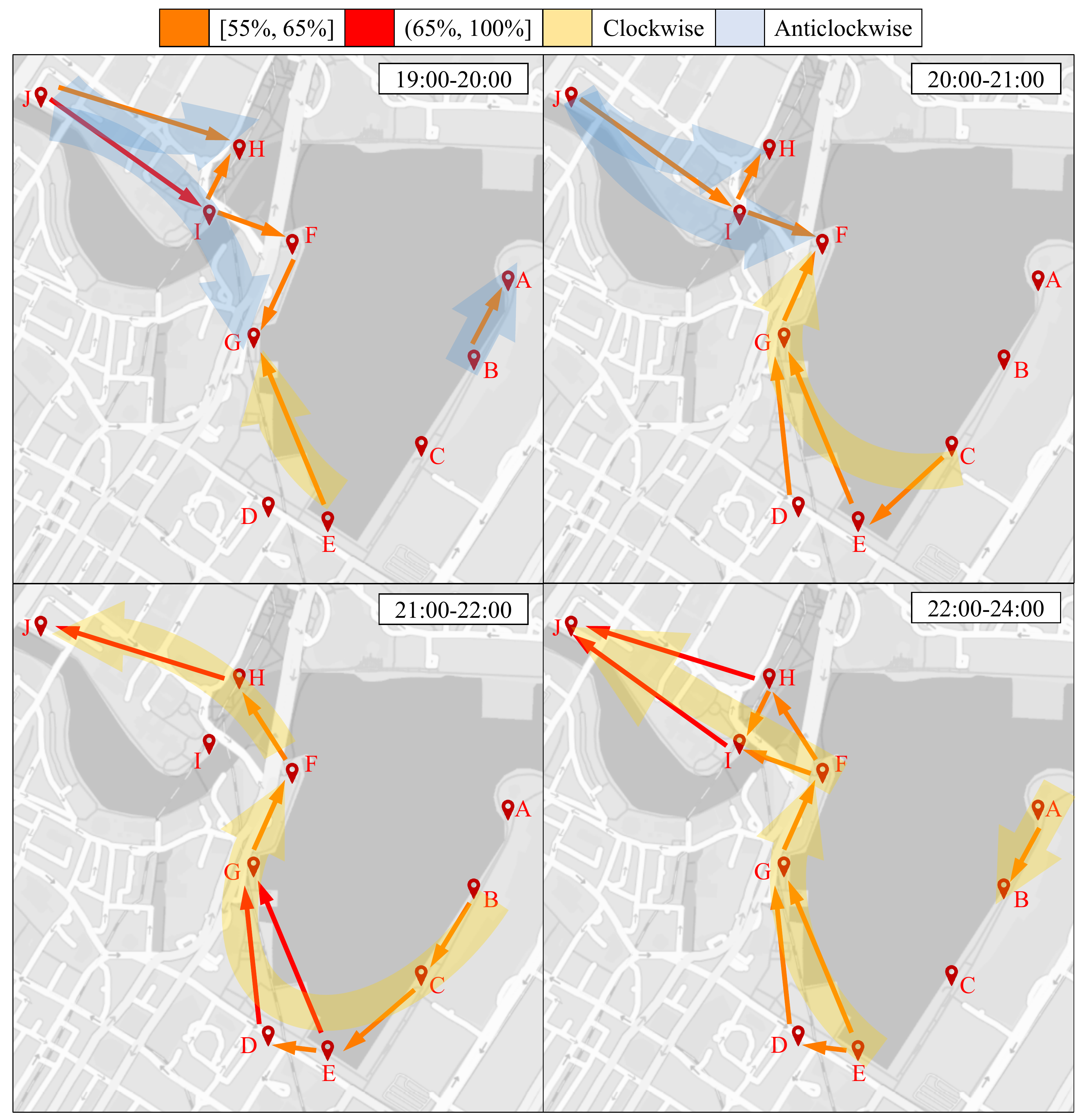}
\caption{Dominant direction (i.e. $|R_t(i,j)-R_t(j,i)| \geq 10\%$) of links of adjacent nodes during different time periods. Percentage is the value of $R_t(i,j)$ computed by Equation~\ref{equa_direct_r}. Two major flow directions are marked, namely clockwise and anticlockwise with respect to the central water area.}
\label{fig_direction_splits}
\end{figure}

The values of $R_t(i,j)$ for links of adjacent nodes are obtained from the above procedure and visualized in Fig.~\ref{fig_direction_splits}, in which only dominant link direction is shown for each link. If a link is in mutual status, it is not shown in the figure. Also, we identify two major flow directions, clockwise and anticlockwise with respect to the central water area, to describe the trend of the dominant link directions. Clockwise flow direction starts from node A and ends at node J along the bay area, while anticlockwise flow direction is the reverse, starting from node J and ending at node A along the bay area.

From the figure, it is clear that there is a change of major flow direction over time. During 7-8 pm, the major crowd flows include both clockwise flow (i.e. from E to G) and anticlockwise flows (i.e. from J to H, from J to G, and from B to A). The clockwise direction was weaker at this time. One hour later, anticlockwise flows became weaker: flow from J to H lost one link (i.e. J-H), flow from J to G was cut to F, and flow from B to A disappeared. At the same time, clockwise flows became stronger, as two dominant link directions, D to G and C to E, took place. Afterward, during 9-10 pm, the anticlockwise flow was completely missing in the area, and the main trend of crowd movement was clockwise, from B to J. Lastly, during 10 pm - 12 am, there was still no anticlockwise flows. Clockwise flows from F to J became stronger by adding two dominant link directions (i.e. F-I and I-J), clockwise flows from B to E were lost, and clockwise flows from E to G became weaker.

Overall, one can see that the crowd flow trend in the area covering node F to J had large changes over time. During sunset and early night, visitors in this area would move from node J through H and I towards F. Whereas, during late-night, they would move from node F backward to J. As for the crowd flows in the area covering node A to G, they showed an overall trend from B/C to G/F through D/E for most of the time, and this trend reached its peak during 9-10 pm.

\section{Conclusion and Future Work}
\label{sec_conclusion}

\subsection{Conclusion}

In this paper, a comprehensive data analysis framework was presented to extract spatial, temporal, and spatiotemporal patterns related to crowd behaviors in a large social event from WiFi probe request records. Three types of clustering methods, namely hierarchical agglomerative clustering, $k$-means clustering, and $k$-shape clustering, along with statistics and visualization were applied to extract those patterns. A real-world dataset collected in a large social event was used to demonstrate the effectiveness of the proposed approaches. Different hidden patterns beneath the data were obtained and discussed, which are usually not available to the event managers and shed light upon the understanding of crowd behaviors in the event.

In Section~\ref{sec_spatial}, three main spatial patterns of crowd movement were obtained:
\begin{enumerate*}
  \item majority of the trajectories are shorter than three nodes,
  \item popularity of each node is highly affected by surrounding POIs,
  \item the event area can be divided into Zone I and Zone II and the transition between Zone I and Zone II is infrequent.
\end{enumerate*}
Based on these findings, if the event managers want to make people visit more spots in the event, more attractive art installations should be placed in spots that are further to the surrounding POIs. Also, they should make the space between node G and node E more attractive such that the Zone I and Zone II can be more connected.
In Section~\ref{sec_temporal}, we discovered four clusters of days based on their overall device count curves. The summarized characteristics of each cluster of days can help the event managers to plan differently for different types of days. In addition, four clusters of nodes were obtained based on their individual device count curves. The characteristics of the four clusters demonstrate to the event managers the positive influences of the scheduled light shows as well as capture the special ending patterns at node A and node B.
In Section~\ref{sec_spatiotemporal}, the change of trajectory duration over the length and the change of crowd movement over time were studied and discussed. The former provides the event managers with information on how much time visitors spent in the event, and the latter summarizes the overall trends of crowd movement during different time periods.

\subsection{Future Work}

The potential future work of this study is threefold.

First, because not every person carries a mobile device that has WiFi turned on, and different mobile devices send out probe requests with different frequencies, the count of MAC addresses detected by a WiFi sensing node is thus correlated but not equal to the count of people passing by the node. To precisely count the visitors from WiFi probe requests, therefore, requires the collection of ground truth (i.e. the real number of people) and the construction of a generalized model that maps the count of MAC addresses to the real number of people. How to effectively and accurately realize these two steps remains a challenge.

Second, since the locally assigned MAC addresses are not globally unique and keep changing, they are not directly suitable for investigating the spatial or spatiotemporal patterns of the crowd. However, simply excluding them from the related analysis could lose valuable information. Therefore, it is an interesting research challenge on how to extract more information on people's movement from these locally assigned MAC addresses.

Lastly, as discussed in Section~\ref{sec_intro}, compared with GPS, passive WiFi sensing does not require the active participation of people, and thus it can save manpower cost on data collection and provide a much larger coverage on the entire population of participants. But at the same time, WiFi sensing also has its limitation in space coverage: there is no precise information about people's movement between two sensing nodes. As a result, designing a comprehensive crowd monitoring system that combines GPS tracking and passive WiFi sensing is another worth-trying direction, which could be a more balanced solution to monitor crowd behaviors in large outdoor social events.

\section*{Acknowledgment}
The authors would like to thank the support of the Singapore Urban Redevelopment Authority (URA) on this work.


\Urlmuskip=0mu plus 1mu
\bibliographystyle{IEEEtran}
\bibliography{main}

%




\begin{IEEEbiography}[{\includegraphics[width=1in,height=1.25in,clip,keepaspectratio]{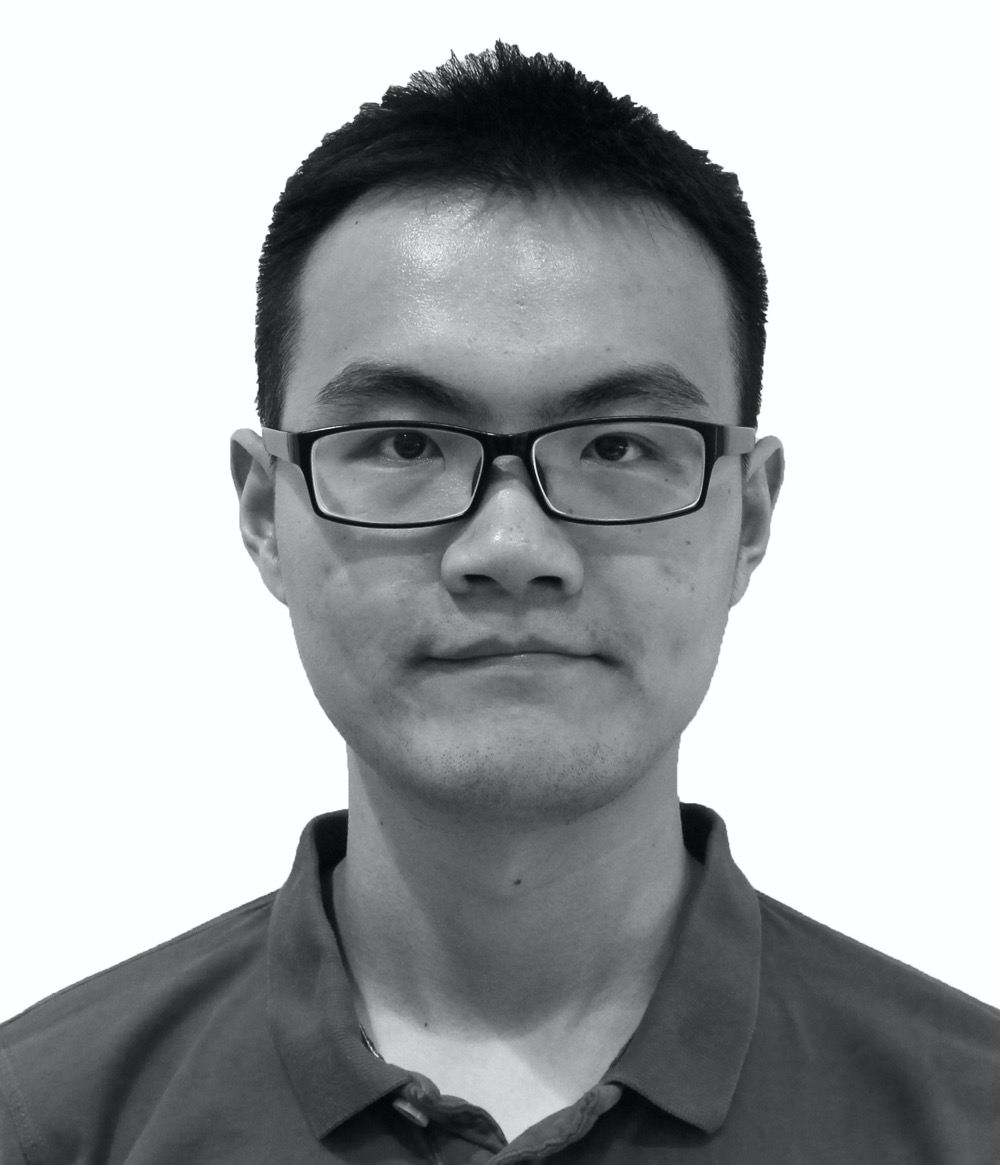}}]{Yuren Zhou}
received the B.Eng. degree in Electrical Engineering from Harbin Institute of Technology, Harbin, China in 2014, and the Ph.D. degree from Singapore University of Technology and Design, Singapore in 2019, with a focus on data mining and smart city applications. He is currently a postdoctoral research fellow at Singapore University of Technology and Design. His current research interests include big data analysis, machine learning, and their applications in urban human mobility, building energy management, and Internet of Things.
\end{IEEEbiography}

\begin{IEEEbiography}[{\includegraphics[width=1in,height=1.25in,clip,keepaspectratio]{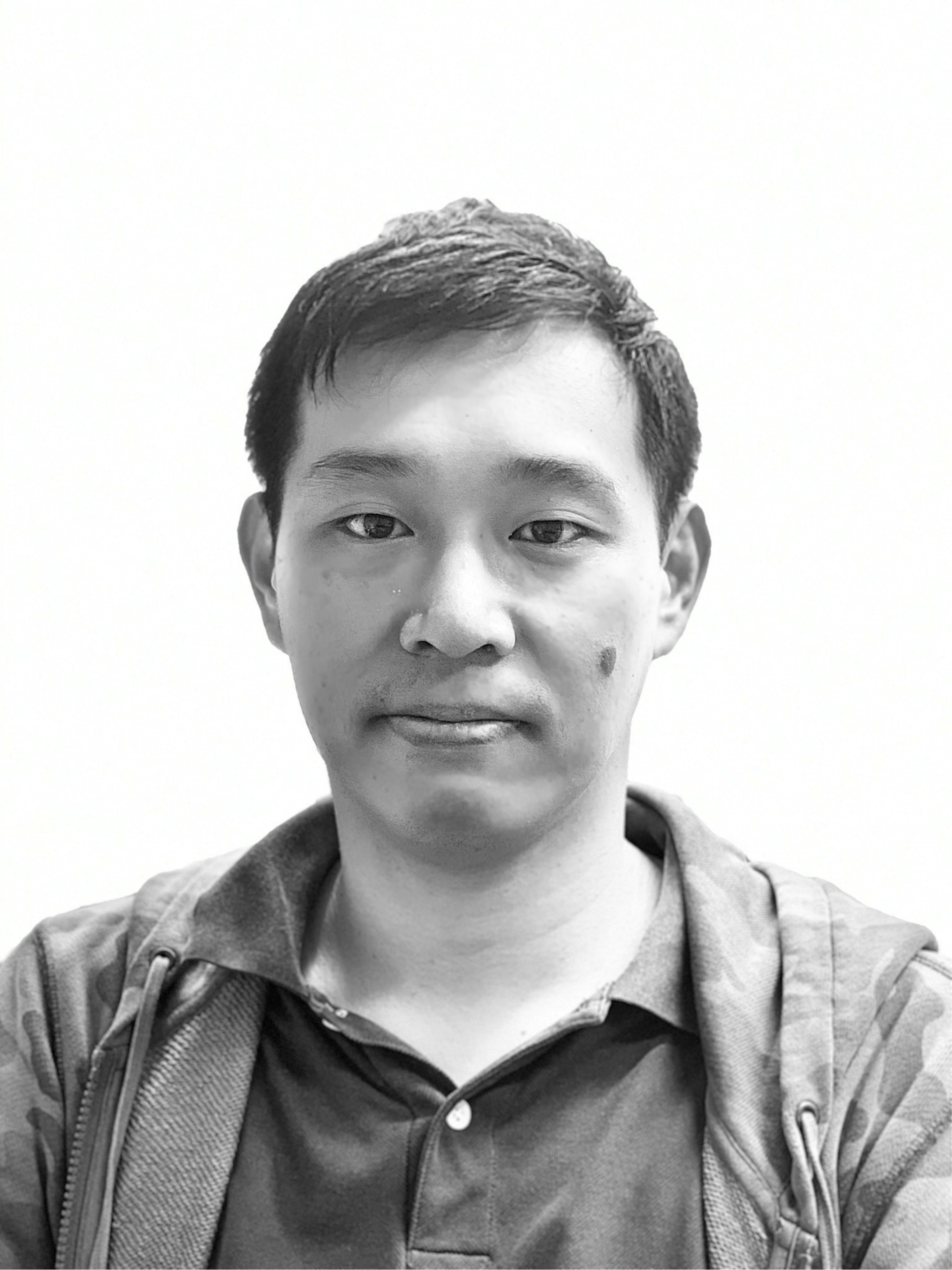}}]{Billy Pik Lik Lau}
received the degree in computer science and M.Phil. degree in computer science from Curtin University, Perth, WA, Australia, in 2010 and 2014, respectively. He is currently a Ph.D. candidate with Dr. Chau Yuen at the Singapore University of Technology and Design, Singapore. He studied the cooperation rate between agents in multiagents systems during master studies. His current research interests include urban science, big data analysis, data knowledge discovery, Internet of Things, and unsupervised machine learning.
\end{IEEEbiography}

\begin{IEEEbiography}[{\includegraphics[width=1in,height=1.25in,clip,keepaspectratio]{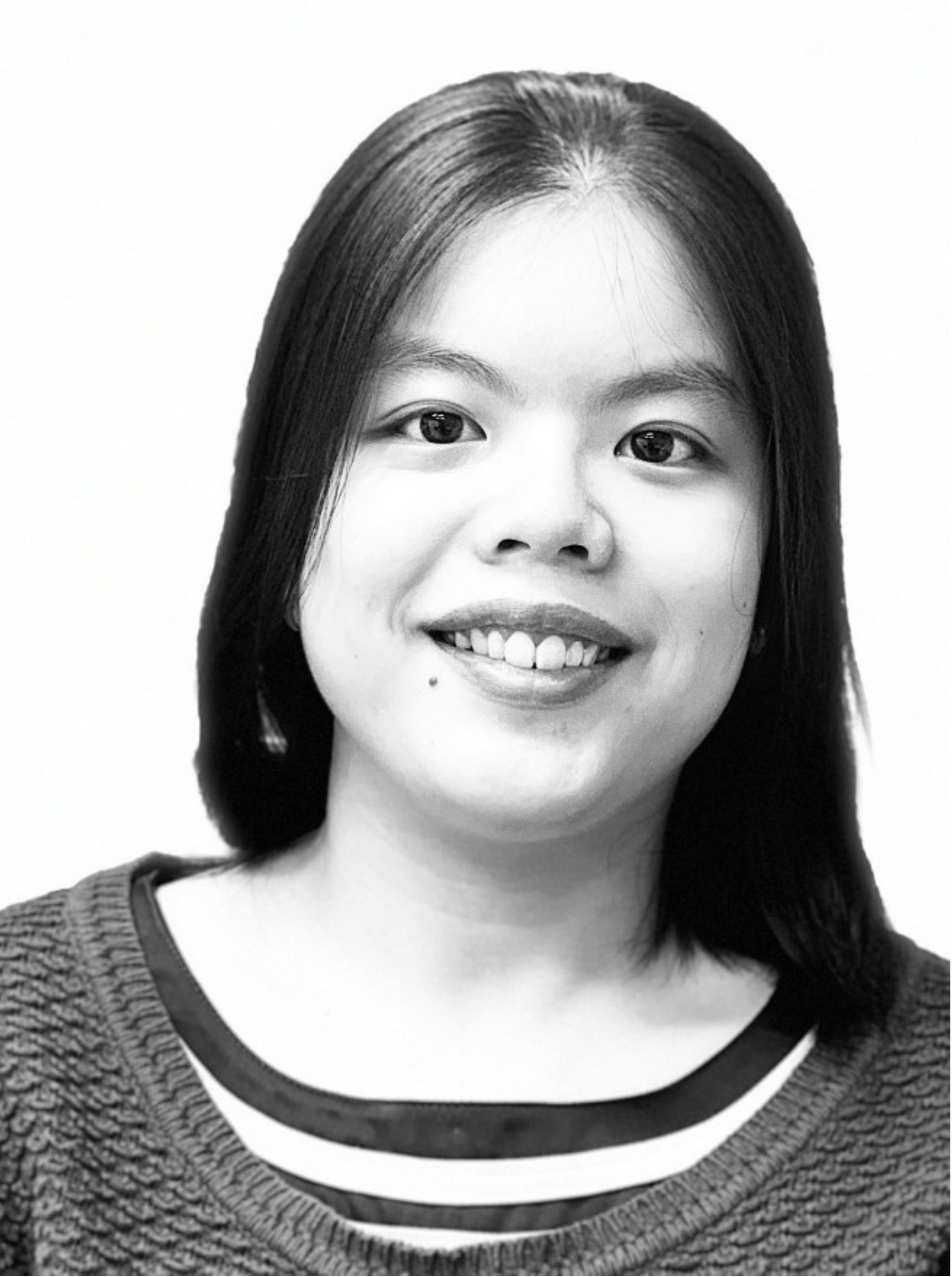}}]{Zann Koh}
received the B.Eng. degree in Engineering and Product Development from the Singapore University of Technology and Design, Singapore, in 2017. She is currently pursuing the Ph.D. degree with the Singapore University of Technology and Design, Singapore, under Dr. Chau Yuen’s supervision. Her current research interests include big data analysis, data discovery, urban human mobility, and unsupervised machine learning.
\end{IEEEbiography}

\begin{IEEEbiography}[{\includegraphics[width=1in,height=1.25in,clip,keepaspectratio]{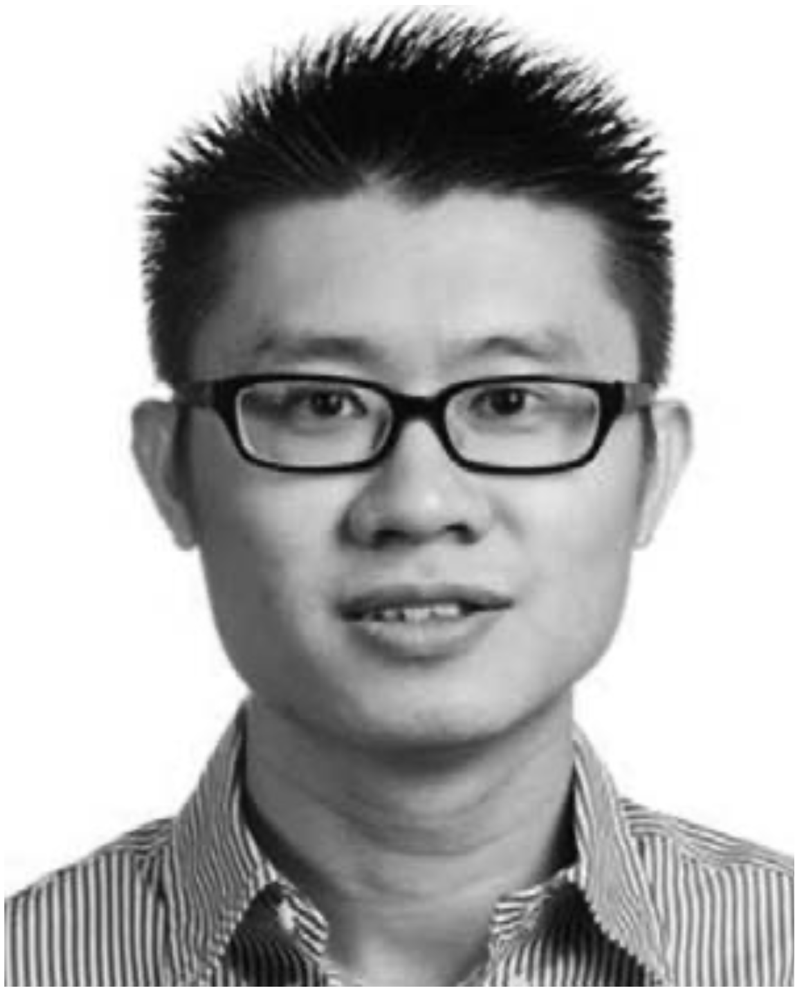}}]{Chau Yuen}
is currently an Associate Professor at Singapore University of Technology and Design. He received the B.Eng. and Ph.D. degrees from Nanyang Technological University, Singapore, in 2000 and 2004, respectively. He was a Postdoctoral Fellow at Lucent Technologies Bell Labs, Murray Hill, NJ, USA, in 2005. He was a Visiting Assistant Professor at The Hong Kong Polytechnic University in 2008. From 2006 to 2010, he was a Senior Research Engineer at the Institute for Infocomm Research (I2R, Singapore), where he was involved in an industrial project on developing an 802.11n Wireless LAN system, and participated actively in 3Gpp Long Term Evolution (LTE) and LTE-Advanced (LTE-A) Standardization. He has been with the Singapore University of Technology and Design since 2010.

Dr. Yuen is a recipient of the Lee Kuan Yew Gold Medal, the Institution of Electrical Engineers Book Prize, the Institute of Engineering of Singapore Gold Medal, the Merck Sharp and Dohme Gold Medal, and twice the recipient of the Hewlett Packard Prize. He received the IEEE Asia-Pacific Outstanding Young Researcher Award in 2012. He serves as an Editor for the IEEE Transaction on Communications and the IEEE Transactions on Vehicular Technology and was awarded the Top Associate Editor from 2009 to 2015.
\end{IEEEbiography}

\begin{IEEEbiography}[{\includegraphics[width=1in,height=1.25in,clip,keepaspectratio]{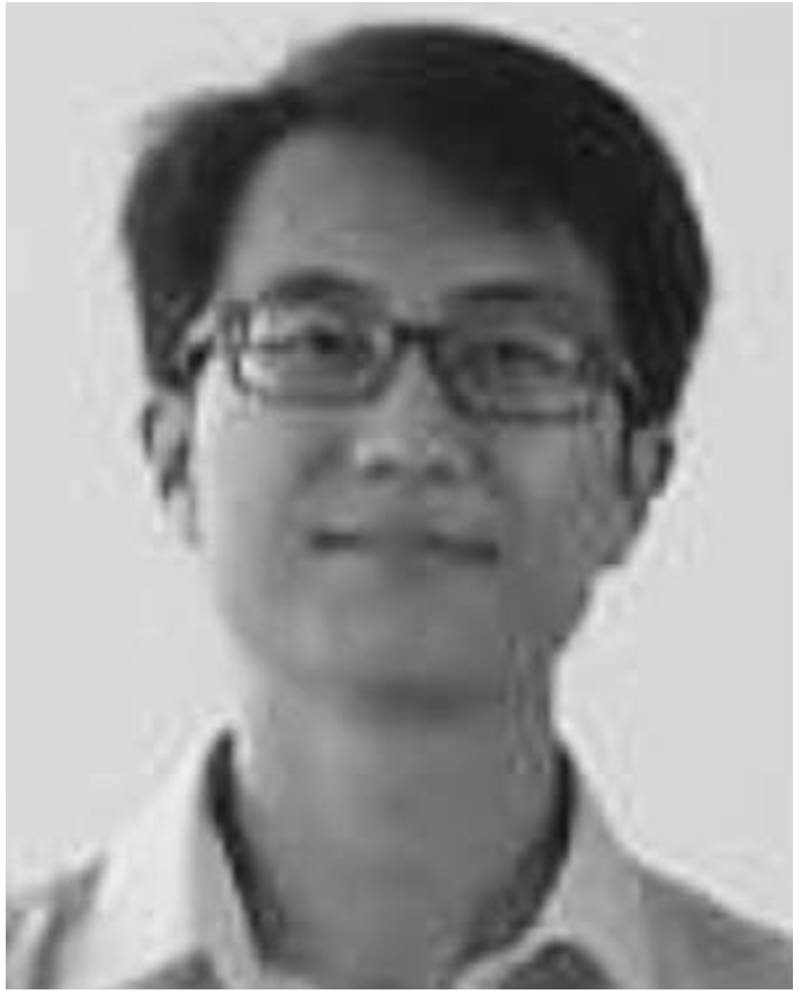}}]{Benny Kai Kiat Ng}
received the degree in electrical and communication engineering from Curtin University, Perth, WA, Australia, in 2013. He has been involved with several research projects with Dr. Chau Yuen. His current research interests include sensors, data collection, signal processing, hardware, and embedded system development.
\end{IEEEbiography}




\end{document}